\documentclass[twocolumn,floatfix,preprintnumbers,nofootinbib]{revtex4-2}
\usepackage[utf8]{inputenc}
\pdfoutput=1
\usepackage[colorlinks=true,citecolor=blue,linkcolor=blue]{hyperref}
\usepackage[normalem]{ulem}
\usepackage{url}
\usepackage{graphicx,wrapfig,float,slashed,cancel}
\usepackage{amsmath,amssymb,epsfig,graphicx,xcolor,stmaryrd}
\usepackage{physics}
\usepackage{bm}
\usepackage{bbm}
\usepackage{enumitem}
\begin{document}
\title{Charming ALPs}
\author{Adrian Carmona}
\email{adrian@ugr.es}
\affiliation{CAFPE and Departamento de F\'isica Teórica y del Cosmos,\\ Universidad de Granada, E18071 Granada, Spain}
\author{Christiane Scherb }
\email{cscherb@uni-mainz.de}
\affiliation{PRISMA$^+$ Cluster of Excellence \& Mainz Institute for Theoretical Physics,\\ Johannes Gutenberg University, 55099 Mainz, Germany}
\author{Pedro Schwaller}
\email{pedro.schwaller@uni-mainz.de}
\affiliation{PRISMA$^+$ Cluster of Excellence \& Mainz Institute for Theoretical Physics,\\ Johannes Gutenberg University, 55099 Mainz, Germany}

\date{\today}
\preprint{MITP-21-003}
\begin{abstract}
	Axion-like particles (ALPs) are ubiquitous in models of new physics explaining some of the most pressing  puzzles of the Standard Model. However, until relatively recently, little attention has been paid to its interplay with flavour. In this work, we study in detail the phenomenology of ALPs that exclusively interact with up-type quarks at the tree-level, which arise in some well-motivated ultra-violet completions such as QCD-like dark sectors or Froggatt-Nielsen type models of flavour. Our study is performed in the low-energy effective theory to highlight the key features of these scenarios in a model independent way. We derive all the existing  constraints on these models and demonstrate how upcoming experiments at fixed-target facilities and the LHC can probe regions of the parameter space which are currently not excluded by cosmological and astrophysical bounds. We also emphasize how a future measurement of the currently unavailable meson decay $D\to \pi +\rm{invisible}$ could complement these upcoming searches.  
\end{abstract}

\maketitle
\section{Introduction}

One of the outstanding open questions in particle physics is the nature of dark matter (DM) and whether it is part of a larger dark sector that we yet have to discover. Most realistic models require some form of non-gravitational interaction between us and the dark sector in order to satisfy cosmological constraints. These "portals" then offer the possibility to probe the dark sector through laboratory experiments or astrophysical observations. 

An intriguing possibility is that the portal to the dark sector is flavour sensitive or even connected to an ultra-violet (UV) theory of flavour. Simple flavoured dark matter scenarios have received a lot of attention recently~\cite{Craig:2015pha, Agrawal:2014aoa, Batell:2011tc, Calibbi:2015sfa}, since they allow probes of dark sector physics using low energy flavour observables. For strongly coupled dark sectors, a flavoured portal imprints the Standard Model (SM) flavour structure on the dark sector, leading to a variety of new phenomena~\cite{Renner:2018fhh, Mies:2020mzw}. 

So far only couplings to down-type quarks were considered in this context, therefore it is natural to ask what new aspects arise if the portal couples to up-type quarks instead.~\footnote{Simple DM models which dominantly couple to up-type quarks were also studied e.g. in~\cite{Blanke:2017tnb,Jubb:2017rhm,Blanke:2020bsf}.} The key feature in this case is the emergence of a light pseudo Nambu-Goldstone boson (pNGB) which dominantly couples to up-type quarks, and which we therefore dub a \textit{charming} ALPs. 

Our main goal in this work is to develop the effective theory of the charming ALP and its phenomenological profile, independently of its embedding in different UV scenarios. Besides QCD-like dark sectors~\cite{Strassler:2006im, Bai:2013xga, Schwaller:2015gea, Schwaller:2015gea, Renner:2018fhh, Cheng:2019yai}, these particles can arise e.g. in specific Froggatt-Nielsen (FN) models~\cite{Froggatt:1978nt} where only right-handed (RH) up-quarks have non-zero charges. The different UV completions provide some guidance for the structure of the effective couplings, which we use to define benchmark scenarios for the charming ALP.

We will study the phenomenology of these different benchmark models through their low-energy effective field theory (EFT), taking into account the effects of QCD confinement for low enough ALP masses. A lot of work has been done in this arena, see e.g~\cite{Jaeckel:2015jla, Brivio:2017ije,  Bellazzini:2017neg, Bauer:2017ris, Knapen:2017ebd,  Bauer:2018uxu, Aloni:2018vki} for the study of ALP collider signatures, \cite{Batell:2009jf, Kamenik:2011vy, Gavela:2019wzg} and~\cite{Bauer:2019gfk, Cornella:2019uxs,  Calibbi:2020jvd}  for the study of flavour-changing neutral currents (FCNCs) in the quark and lepton sector, respectively, as well as~\cite{Choi:2017gpf, MartinCamalich:2020dfe, Chala:2020wvs, Bauer:2020jbp} for the calculation of the one-loop running. It is worth to mention also~\cite{Marciano:2016yhf, DiLuzio:2020oah} regarding CP-violating probes of ALPs.

The work is organized as follows: We introduce the effective Lagrangian describing the charming ALPs and motivate briefly the four particular benchmarks models studied in this work in section~\ref{sec:eft}. In section~\ref{sec:flavour} we examine the different flavour constraints relevant for charming ALPs. More specifically, we consider  $D-\bar{D}$ mixing, exotic decays of $D$, $B$ and $K$ mesons as well as the decay $J/\psi\to a\gamma$.  We discuss the bounds arising from astrophysical observables and cosmology in section~\ref{sec:cosm}, describing also the different ALP decay channels. In section~\ref{sec:coll} we study the different collider probes on the models at hand, including  upcoming fix-target experiments as well as those at LHC forward detectors. We combine all these different bounds and discuss the resulting constraints on the parameter space of the models in section~\ref{sec:res}. We conclude in section~\ref{sec:conc}. Finally, we present in some detail  the particular UV completions considered in this work in appendices~\ref{app:dqcd} and~\ref{app:fn}.

\section{Charming ALP EFT}
\label{sec:eft}
We consider a general ALP, which we will denote $a$, with flavour-violating couplings to RH up-quarks. The most general EFT describing such a system is given by the following Lagrangian~\cite{Georgi:1986df, Choi:1986zw}
\begin{widetext}
\begin{align}
	\mathcal{L}&=\frac{1}{2}(\partial_{\mu}a)(\partial^{\mu}a)-\frac{m_a^2}{2}a^2+\frac{\partial_{\mu} a}{f_a} \left[(c_{u_R})_{ij}\bar{u}_{Ri} \gamma^{\mu} u_{Rj}+c_{H}H^{\dagger} i\overleftrightarrow{D_{\mu}} H\right]\nonumber\\
	&-\frac{a}{f_a}\left[c_{g}\frac{g_3^2}{32\pi^2}G_{\mu\nu}^{a}\tilde{G}^{\mu\nu a}+c_{W}\frac{g_2^2}{32\pi^2}W_{\mu\nu}^I\tilde{W}^{\mu\nu I} +c_{B}\frac{g_1^2}{32\pi^2}B_{\mu\nu}\tilde{B}^{\mu\nu}\right],
	\label{eq:alag}
\end{align}
\end{widetext}
where $g_1,g_2$ and $g_3$ are the gauge couplings of $U(1)_Y$, $SU(2)_L$ and $SU(3)$, respectively, whereas $B_{\mu\nu}$, $W_{\mu\nu}^I,\, I=1,2,3,$ and $G_{\mu\nu}^a,\, a=1,\ldots,8,$ are their corresponding field-strength tensors. Furthermore, $\tilde{B}_{\mu\nu}=\frac{1}{2}\varepsilon_{\mu\nu\alpha\beta}B^{\alpha\beta},\ldots,$ denote their corresponding  duals, while $H$ stands for the SM Higgs doublet. The Wilson coefficients (WCs) $c_{g}, c_{W}, c_{B}$ and $c_H\in\mathbb{R}$, whereas $c_{u_R}$ is a hermitian matrix. In order to write down the above Lagrangian we have assumed that $a$ is the pNGB of the spontaneous breaking of some global $U(1)$ symmetry, which is softly broken and  may be anomalous. We have also assumed that the couplings to leptons, SM quark doublets and RH down-type quarks vanish. Here, contrary to the QCD axion case, we will treat $m_a$ and $f_{a}$ as independent parameters. 

Using field redefinitions, we can trade the operator $\mathcal{O}_{H}=(\partial^{\mu}a/f_a) H^{\dagger}i \overleftrightarrow{D_{\mu}} H$ by the flavour-blind and chirality conserving one (see e.g.~\cite{Georgi:1986df,Brivio:2017ije})
\begin{align}
	&\frac{\partial_{\mu} a}{f_a}\left[\frac{1}{3}\bar{q}_{Li}\gamma^{\mu}q_{Li} +\frac{4}{3}\bar{u}_{Ri}\gamma^{\mu}u_{Ri}-\frac{2}{3}\bar{d}_{Ri}\gamma^{\mu}d_{Ri}\right.\nonumber\\
	&\left.-\bar{l}_{Li}\gamma^{\mu}l_{L i}-2\bar{e}_{Ri}\gamma^{\mu}e_{Ri}\right].
\end{align}
   Together with 
   \begin{align}
	   \mathcal{O}_{W}=\frac{a}{f_a}\frac{g_2^2}{32\pi^2}W_{\mu\nu}^I\tilde{W}^{\mu\nu I}
   \end{align}
  this operator induces flavour-changing neutral currents (FCNCs) at one-loop, which have been studied in~\cite{Gavela:2019wzg} in the framework of $B$ and $K$-meson decays. Here we will just assume that both WCs are small enough so that the leading flavour-violating effects are parametrized by $c_{u_R}$. Furthermore, after integrating by parts and using equations of motion, one can express this chirality-conserving operator as a function of 
\begin{align}
	-i\frac{a}{f_a}\bar{q}_{Lk}\tilde{H}  u_{Rj}	\left(Y_{u}\right)_{ks} (c_{u_R})_{sj}+\mathrm{h.c.}
	\label{eq:yuks}
\end{align}
plus some extra contributions to the anomalous terms in~(\ref{eq:alag}), where $\tilde{H}=i\sigma^2 H^{\ast}$ and $Y_u$ is the up Yukawa matrix, 
\begin{align}
	-\bar{q}_{Lk}\tilde{H}  u_{Rj}	\left(Y_{u}\right)_{kj}+\mathrm{h.c.}\,.
\end{align}
Note that, without any loss of generality, we can always choose a basis where 
\begin{align}
	Y_u=\lambda_u,\qquad Y_d=\tilde{V}\lambda_d,
\end{align}
with $\lambda_{u,d}$ diagonal matrices with real and positive entries and $\tilde{V}$ a unitary matrix. In the case where there is no extra contribution to the fermion masses, $\tilde{V}$ is just the CKM mixing matrix $V$ and $\lambda_u=\sqrt{2}\mathcal{M}_u/v$, $\lambda_d=\sqrt{2} \mathcal{M}_d/v$, where $\mathcal{M}_u=\mathrm{diag}(m_u,m_c,m_t)$,  $\mathcal{M}_d=\mathrm{diag}(m_d,m_s,m_b)$, and $v= 246$\,GeV the Higgs vacuum expectation value. In this basis, the RH up-quarks do not need to be rotated to diagonalize the mass matrices generated after electroweak symmetry breaking. Indeed, one could just take
\begin{align}
	U_L^d=\tilde{V}, \quad U_R^d=U_L^u=U_R^u=\mathbbm{1}.
\end{align}
Henceforth, we will assume that the above EFT Lagrangian is defined in such a basis. Then, if we denote the WC of the operators in \eqref{eq:yuks}  by $\mathcal{C}$, one has that 
\begin{align}
	\mathcal{C}_{ij}=(\lambda_u)_{ii} (c_{u_R})_{ij}.
\end{align}

For small ALP masses, $m_a\lesssim 1$~GeV, $a$ will mostly decay to hadrons. These decays will proceed through the following Lagrangian~\cite{Georgi:1986df, Choi:1986zw, Bardeen:1986yb, Krauss:1986bq}
\begin{widetext}
\begin{align}
	\mathcal{L}_{\rm aChPT}&=\frac{1}{2}(\partial_{\mu}a)(\partial^{\mu}a)-\frac{m_{a}^2}{2}a^2 -\frac{a}{f_a}\frac{e^2}{32\pi^2}c_{\gamma}F_{\mu\nu}\tilde{F}^{\mu\nu}+\frac{f_{\pi}^2}{4}\mathrm{Tr}\left(\partial_{\mu}U \partial^{\mu}U^{\dagger}\right)\nonumber\\
	&+\frac{f_{\pi}^2 B_{0}}{2}\mathrm{Tr}\left(\hat{m}_{q}(a)U^{\dagger}+U\hat{m}_q^{\dagger}(a)\right)+i\frac{f_{\pi}^2}{2}\frac{\partial_{\mu}a}{f_a}\mathrm{Tr}\left[ (\hat{c}+\varkappa_q c_{g}) \left(U D^{\mu} U^{\dagger}\right)\right]
	\label{lag:alp},
\end{align}
\end{widetext}
where $B_0$ is a constant, $f_{\pi}\approx 93\,\mathrm{MeV}$ is the pion decay constant and $m_q$ is the quark mass matrix $m_q=\mathrm{diag}(m_u,m_d,m_s)$. In the above equation, 
\begin{align}
	U(\Pi)=\mathrm{exp}\left(2i \Pi/f_{\pi}\right),
\end{align}
where 
\begin{align}
	\Pi=\varphi^a\frac{\lambda^a}{2}=\frac{1}{\sqrt{2}}\begin{pmatrix}\frac{1}{\sqrt{2}}\pi^0+\frac{\eta_{8}}{\sqrt{6}}& \pi^+&K^+\\ \pi^-&-\frac{1}{\sqrt{2}}\pi^0+\frac{\eta_{8}}{\sqrt{6}}&K^0\\ K^-&K^{0}& -\frac{2}{\sqrt{6}}\eta_{8}\end{pmatrix}
\end{align}
is the Goldstone matrix describing the spontaneous symmetry breaking  $SU(3)_L\otimes SU(3)_R\to SU(3)_V$ of QCD.
On the other hand, $\varkappa_q=m_q^{-1}/\mathrm{Tr}(m_q^{-1})$, 	$\hat{c}=\mathrm{diag}((c_{u_R})_{11},0,0)$, and
\begin{align}
	\hat{m}_q(a)&=\mathrm{exp}\left(-i\varkappa_q c_{g}\frac{a}{2 f_a}\right)m_q \left(-i\varkappa_q  c_{g}\frac{a}{2 f_a}\right), \\
	D_{\mu}U&=\partial_{\mu}U+ie A_{\mu}\Big[Q_q,U\Big],\\
	c_{\gamma}&=c_{W}+c_{B} -2\,N_c c_{g} \mathrm{Tr}\left(\varkappa_q Q_q^2\right),
		\end{align}
		with $e=g_2g_1/\sqrt{g_1^2+g_2^2}$ the electric charge and $Q_q=1/3\,\mathrm{diag}(2,-1,-1)$. One should note that, in order to get the above Lagrangian, we had to get rid of the gluon coupling by the following chiral transformation
\begin{align}
	q\to \mathrm{exp}\Big(-i\frac{a}{2f_a}c_{g}\varkappa_q(1+\gamma_5)\Big) q.
\end{align}
Note that this Lagrangian gives an irreducible contribution to the ALP mass
\begin{align}
	m_{a\, \textrm{QCD}}^2&=c_{g}\frac{ m_{\pi}^2 f_{\pi}^2}{(m_d+m_u)f_a^2}\frac{m_u m_d m_s}{m_um_d+m_u m_s + m_d m_s}\nonumber\\
	&+\mathcal{O}\left(\frac{m_{\pi}^2 f_{\pi}^4}{f_a^4}\right),
\end{align}
where
\begin{align}
	m_{\pi}^2= B_0 (m_u+m_d)+\mathcal{O}\left(\frac{m_{\pi}^2f_{\pi}^4}{f_a^4}\right).
\end{align}

Kinetic mixing arising from the last term in eq.~(\ref{lag:alp}) induces a mass mixing between the different neutral  pions. In particular, we obtain
\begin{align}
	\pi&\to \pi-\frac{f_{\pi}}{f_a}\frac{m_a^2}{m_a^2-m_{\pi}^2}\left(\mathcal{K}_{\pi}-\frac{\mathcal{K}_{\eta}\delta_I m_{\pi}^2}{\sqrt{3}(m_a^2-m_{\eta}^2)}\right)a\nonumber\\
	&-\frac{\delta_I m_{\pi}^2}{\sqrt{3}(	m_{\eta}^2-m_{\pi}^2)}\eta_8+\mathcal{O}(f_{\pi}^2/f_a^2)+\mathcal{O}(\delta_I^2),
	\label{eq:mmix}
\end{align}
where
\begin{align}
	\delta_I=\frac{m_d-m_u}{m_d+m_u}\approx \frac{1}{3},\qquad m_{\eta}^2=\frac{m_d+m_u+4m_s}{3(m_u+m_d)}m_{\pi}^2
\end{align}
and
\begin{align}
	\mathcal{K}_{\pi}&=c_{g}\frac{m_s(m_d-m_u)}{2(m_sm_u+m_dm_u+m_dm_s)}+\frac{(c_{u_R})_{11}}{2},\\
	\mathcal{K}_{\eta}&=c_{g}\frac{m_s(m_d+m_u)-2m_dm_u}{2\sqrt{3}((m_sm_u+m_dm_u+m_dm_s))}+\frac{(c_{u_R})_{11}}{2\sqrt{3}}.
\end{align}

As already mentioned, we are interested in scenarios where the ALP only interacts with RH up-quarks, and it is likely to mediate flavour-changing processes. In order to explore how the different experimental constraints are intertwined and the best way to probe these models, we will consider four different benchmarks. In practice, each of these scenarios corresponds  to a particular choice of the matrix $c_{u_R}$ defined above. 

The first two benchmarks are motivated by theories of 'dark QCD', where the SM is extended with a confining dark sector, composed of $n_d$ dark flavours transforming under $SU(N_d)$. These models constitute a particular UV completion of the scenarios we have in mind, where light pNGB bosons do only interact at leading order with the SM RH up-quarks. Indeed, this is a natural outcome when both sectors are mediated by a scalar, bifundamental of both confining groups,  with hypercharge $-2/3$. We refer the reader to appendix~\ref{app:dqcd} and references therein for more details. At the end of the day, after integrating out the heavy scalar mediator, and assuming confinement in the dark QCD group, we obtain almost degenerate, parametrically light scalars  with a Lagrangian along the lines of	\eqref{eq:alag}. When studying phenomena where the interplay of the different scalars is not relevant, one can examine the phenomenological impact of the different degrees of freedom separately. In particular, focusing on the 'diagonal' dark pions $\pi_{D_3}$ and $\pi_{D_8}$, we obtain 
\begin{align}
	c_{u_R}^{(3)}&=\frac{-\kappa_0^2}{4}\begin{pmatrix}9c_{12}^2-4s_{12}^2& 13 c_{12}s_{12}&0\\13c_{12}s_{12} & 9s_{12}^2-4c_{12}^2  &0\\ 0&0&0\end{pmatrix},\\
		c_{u_R}^{(8)}&=\frac{-\kappa_0^2}{4\sqrt{3}}\begin{pmatrix}4s_{12}^2+9c_{12}^2& 5c_{12}s_{12}&0\\5c_{12}s_{12} & 4c_{12}^2+9s_{12}^2 &0\\ 0&0&-2\end{pmatrix},
\end{align}
as well as $c_H=0$ and $c_{g}=c_W=c_B=0$ at tree level. In the equation above $\kappa_0\in \mathbb{R}^+$ and $c_{12}=\cos\theta_{12}$, $s_{12}=\sin\theta_{12}$, with $\theta_{12}\in[0,\pi]$. For the sake of concreteness, following~\cite{Renner:2018fhh}, we fix $\theta_{12}=0.022$. 

There is another class of models that can naturally UV complete these scenarios. It is the case of FN models where only RH up-quarks have non-zero charges. Such models are attractive because they naturally result in enhanced Yukawa couplings, while still being in agreement with existing flavour bounds. They where already  considered in a slightly different context \cite{Bauer:2015fxa, Bauer:2015kzy, Bauer:2017cov} but without paying attention to the phenomenology of the light scalar degree of freedom, \emph{the flavon}. We refer the reader to appendix~\ref{app:fn} and references therein for more details. One can reproduce the hierarchical up-quark masses with RH up-quark charges $n_u=(2,1,0)$ under the $U(1)$ flavour group, assuming that the vev of the scalar breaking such symmetry is $\epsilon\sim m_c/m_t$ times its UV cutoff.  At the end of the day, this setup leads to an ALP Lagrangian along the lines of equation~\eqref{eq:alag} with
\begin{align}
	c_{u_R}\sim\begin{pmatrix}2&3\epsilon&3\epsilon^2\\ 3\epsilon&1&\epsilon\\3 \epsilon^2& \epsilon & \epsilon^2\end{pmatrix},
\end{align}
whereas $c_H=0$. The concrete values of the anomalous couplings $c_g, c_W, c_B$ depend on the specific UV completion of the FN model and may all be zero, which is the case we will consider in the following. We define  the FN-motivated benchmark with $c_{u_R}$ given as above, whereas the other WCs are zero.

Finally, we  also consider an infra-red (IR) motivated scenario, where $(c_{u_R})=1$, $\forall i,j$ at the scale $f_a$. This is representative of the anarchic limit, where no hierarchies are present in $c_{u_R}$ and all the entries are of the same order. Similarly to the previous cases, we assume that the anomalous gauge couplings and $c_H$ are negligible.  
 
 \section{Flavour constraints}
 \label{sec:flavour}

\begin{figure}[h]
	\begin{center}
		\includegraphics[width=0.47\linewidth]{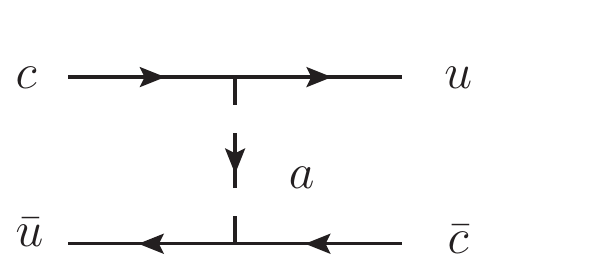}	
		\quad
		\includegraphics[width=0.47\linewidth]{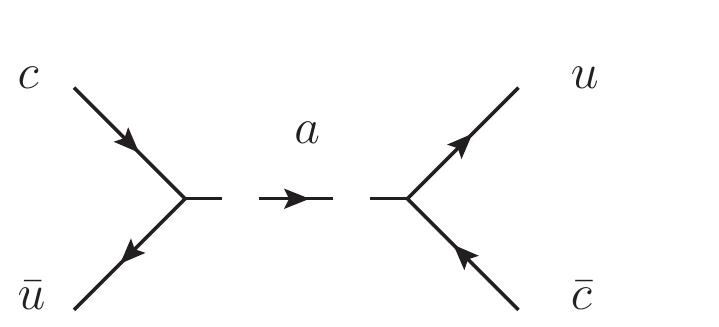}	
		\caption{Parton level diagrams for ALP-mediated $D-\bar{D}$ mixing.}
		\label{fig:mixing}
	\end{center}
\end{figure}

\begin{figure}[h]
	\begin{center}
		\includegraphics[width=0.475\linewidth]{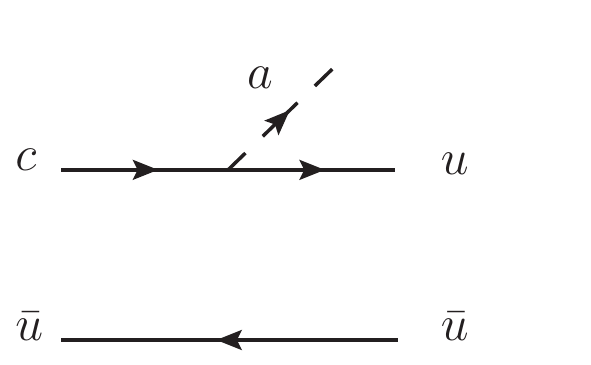}
		\includegraphics[width=0.475\linewidth]{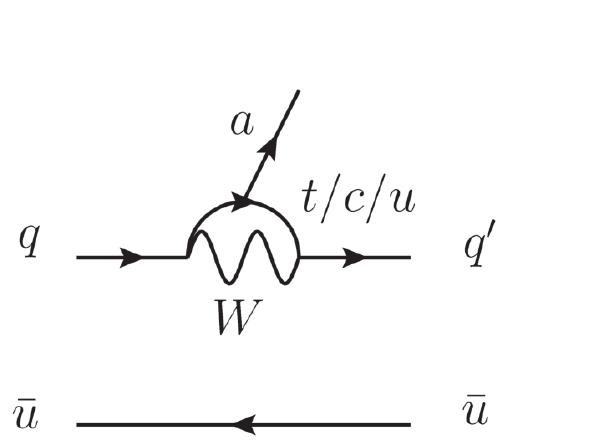}
		\caption{Parton level diagrams for exotic $D$, $K$ and $B$ meson decays involving ALPs.}
		\label{fig:decays}
	\end{center}
\end{figure}

 The presence of  flavour-changing ALP couplings to RH up-quarks will induce several flavour violating processes, constraining significantly the parameter space.  In particular, we will consider
 \begin{itemize}
 	\item the $\Delta F=2$ process of $D-\bar{D}$ mixing displayed in fig.~\ref{fig:mixing} and
 	\item $\Delta F=1$ processes like the exotic decays of $D$, $B$ and $K$ mesons (see fig.~\ref{fig:decays}). 
 \end{itemize}  
 As shown in fig.~\ref{fig:decays}, in the models at hand, exotic $D$ meson decays are tree level processes while $B$ and $K$ decays can only happen at one loop. 
In addition, ALPs also contribute to radiative $J/\psi$ decays (c.f. fig.~\ref{fig:jpsi}). 
 
\subsection{$D-\bar{D}$ mixing}

The effective Hamiltonian relevant for $D$ meson mixing reads~\cite{Ciuchini:1998ix}
\begin{align}
	\mathcal{H}_{\rm eff}^{\Delta C=2}=\sum_{i=1}^5C_i \mathcal{O}_i+\sum_{i=1}^3 \tilde{C}_i\tilde{\mathcal{O}}_i,
\end{align}
where
\begin{align}
	\mathcal{O}_1&=(\bar{c}_L^{\alpha} \gamma^{\mu} u_L^{\alpha})(\bar{c}_L^{\beta} \gamma_{\mu} u_L^{\beta}),\\
	\mathcal{O}_2&=(\bar{c}_R^{\alpha} u_L^{\alpha})(\bar{c}_R^{\beta} u_L^{\beta}),\quad \mathcal{O}_3=(\bar{c}_R^{\alpha} u_L^{\beta})(\bar{c}_R^{\beta} u_L^{\alpha}),\\
	\mathcal{O}_4&=(\bar{c}_R^{\alpha} u_L^{\alpha})(\bar{c}_L^{\beta} u_R^{\beta}),\quad \mathcal{O}_5=(\bar{c}_R^{\alpha} u_L^{\beta})(\bar{c}_L^{\beta} u_R^{\alpha}),
\end{align}
and $\tilde{\mathcal{O}}_{1,2,3}$ are obtained from $\mathcal{O}_i$ after exchanging both chiralities, i.e., $L\leftrightarrow R$. Henceforth, we will just be focusing on the new physics contribution to these WCs, i.e., $C_i=C_i^{\rm NP}$ and $\tilde{C}_i=\tilde{C}_i^{\rm NP}$. Depending on the specific mass of the ALP, such contributions will involve either short or long-distance physics. The first case occurs when integrating out a heavy enough ALP,  $m_a\gg m_c$, whereas the second one is the consequence of applying naively the operator product expansion (OPE) in powers of $\sim 1/m_c$ to the $D-\bar{D}$ system, when $m_a\ll m_c$. In the first case, one obtains
\begin{align}
	\tilde{C}_2&=\frac{(c_{u_R})_{21}^2}{2m_a^2}\frac{m_c^2}{f_a^2},\ C_2=\tilde{C}_2\frac{m_u^2}{m_c^2},\ C_4=-2\tilde{C_2}\frac{m_u}{m_c},
\end{align}
and zero elsewhere, while in the second one 
\begin{align}
	\tilde{C}_2&=-\frac{(c_{u_R})_{21}^2}{2f_a^2},\ C_2=\tilde{C}_2\frac{m_u^2}{m_c^2},\ C_4=-2\tilde{C_2}\frac{m_u}{m_c},
\end{align}
with all other WCs vanishing. In general,
\begin{align}
	2 m_D M_{12}^{\rm NP}&=\sum_{i=1}^5 C_i(\mu)\langle D^0 | \mathcal{O}_i | \bar{D}^0\rangle (\mu)\nonumber\\
	&+\sum_{i=1}^3\tilde{C}_i(\mu)\langle D^0 | \tilde{\mathcal{O}}_i | \bar{D}^0\rangle (\mu),
\end{align}
where $\langle D^0 | \tilde{\mathcal{O}}_i | \bar{D}^0\rangle = \langle D^0 |\mathcal{O}_i | \bar{D}^0\rangle$, for $i=1,2,3$, due to parity conservation of QCD and $m_{D}=1.865\,$GeV~\cite{Zyla:2020zbs}. For small ALP masses, we get~\cite{Bazavov:2017weg}
\begin{align}
	|M_{12}^{\rm NP}|=\frac{1}{2 m_D} \frac{(c_{u_R})_{21}^2}{2f_a^2}\, (0.1561\, \mathrm{GeV}^4),
\end{align}
where $\langle \mathcal{O}_2\rangle =-0.1561\,\mathrm{GeV}^4$ at $\mu=3$\,GeV, see~\cite{Bazavov:2017weg}, and we have neglected $\mathcal{O}(m_u/m_c)$ corrections. However, when studying short-distance physics, it is also necessary to run the different WCs from the scale of integration $\Lambda=m_a$, to the scale $\mu\sim m_c\sim 3\,$GeV. Neglecting $\mathcal{O}(m_u/m_c)$ effects at the UV, we obtain~\cite{Golowich:2009ii}
\begin{align}
	\tilde{C}_2(\mu)&=\left[ r(\mu,\Lambda)^{\frac{1-\sqrt{241}}{6}}\left(\frac{1}{2}-\frac{52}{\sqrt{241}}\right)\right.\nonumber\\
	&\left.+\,r(\mu,\Lambda)^{\frac{1+\sqrt{241}}{6}}\left(\frac{1}{2}+\frac{52}{\sqrt{241}}\right)\right]\tilde{C}_2(\Lambda),\\
	\tilde{C}_3(\mu)&=\frac{705}{32\sqrt{241}}\left[r(\mu,\Lambda)^{\frac{1-\sqrt{241}}{6}}-r(\mu,\Lambda)^{\frac{1+\sqrt{241}}{6}}\right]\tilde{C}_2(\Lambda),\nonumber
\end{align}
where
\begin{align}
	r(\mu,\Lambda)=\left(\frac{\alpha_s(\Lambda)}{\alpha_s(m_t)}\right)^{2/7}\left(\frac{\alpha_s(m_t)}{\alpha_s(m_b)}\right)^{6/23}\left(\frac{\alpha_s(m_b)}{\alpha_s(\mu)}\right)^{6/25}.
\end{align}
Due to the running, to evaluate $M_{12}$ we also need $\langle \mathcal{O}_3\rangle$ at $\mu=3\,$GeV, which reads $0.0464\,\mathrm{GeV}^4$~\cite{Bazavov:2017weg}. As an example, for $\Lambda=m_a=2\,$TeV and $\mu=3\,$GeV, we obtain
\begin{align}
	|M_{12}^{\rm NP}|=\frac{1}{2 m_D} \frac{(c_{u_R})_{21}^2}{2f_a^2}\, (8.95\cdot 10^{-9}\, \mathrm{GeV}^4).
\end{align}

We demand that the new physics contribution to $x_{12}=2|M_{12}|/\Gamma$  does not exceed its upper bound at 95\% confidence level (CL)~\cite{Amhis:2019ckw}, i.e., 
\begin{align}
	x_{12}^{\rm NP}=\frac{2|M_{12}^{\rm NP}|}{\Gamma_D}<0.63\cdot 10^{-2},
\end{align}
where $\Gamma_D=1.60497\cdot 10^{-12}\,\mathrm{GeV}$~\cite{Zyla:2020zbs}.

\subsection{Exotic D, B and K decays}
We study $\Delta F = 1$ decays of the form $M\to N a$ with $M = D^{\pm,0}, B^{\pm,0}, K^{\pm,0}$ and $N = \pi^{\pm,0} , K^{\pm,0}$. The corresponding parton level Feynman diagram are shown in fig.~\ref{fig:decays}. The associated matrix element can be decomposed as
\begin{align}
	\langle N(p^{\prime})| \bar{q}_{i}\gamma_{\mu} q_j| M(p)\rangle=(p+p^{\prime})_{\mu} f_+^{MN}(k^2)+k_{\mu} f_-^{MN}(k^2)\nonumber\\
	 \end{align}
with $k_{\mu}=(p-p^{\prime})_{\mu}$ the momentum transfer and $q_i$ and $q_j$ the relevant quarks for the decay at the parton level. The scalar form factor is then defined as
\begin{align}
	f_0^{MN}(k^2)=f_{+}^{MN}(k^2)+\frac{k^2}{m_M^2-m_{N}^2}f_-^{MN}(k^2)
\end{align}
and the resulting decay width is given by
\begin{align}
	&\Gamma(M \to N a)=\frac{m_M^3|\varkappa_{MN}|^2}{64\pi f_a^2}\left(1-\frac{m_{N}^2}{m_M^2}\right)^2 (f_0^{MN}(m_a^2))^2\nonumber\\ 
	&\times\sqrt{\left(1-\frac{(m_{N}+m_a)^2}{m_M^2}\right)\left(1-\frac{(m_{N}-m_a)^2}{m_{M}^2}\right)},
	\label{eq:master}
\end{align}
where $\varkappa_{MN}$ is defined by
\begin{align}
	\mathcal{L}\supset \varkappa_{MN} \frac{\partial^{\mu}a}{2 f_a} \bar{q}_i \gamma_{\mu} q_j+\mathrm{h.c.}.
\end{align}

In the model at hand and neglecting small isospin-breaking effects, the exotic $D$ meson decay $D^{\pm,0} \to \pi^{\pm,0} a$  is induced by the dimension-5 operator
\begin{align}
	\mathcal{L}\supset \left(c_{u_R}\right)_{ij}\frac{\partial_{\mu} a}{f_a}\left(\bar{u}_R^i\gamma^\mu u_R^j\right).
\end{align}
The width for such decay channel can be  read from equation~\eqref{eq:master} by simply replacing $\varkappa_{MN}$ with $(c_{u_R})_{12}$, $m_M=m_{D}$, $m_{N}=m_{\pi}$ and using $f_{0}^{D\pi}(m_a^2)$ from~\cite{Lubicz:2017syv}.

On the other hand, the one-loop running of $c_{u_R}$ from the UV scale $f_a$ to the IR scale $\mu$ will generate a term~\footnote{The one-loop running also generates a non-zero WC for $\mathcal{O}_H$,  but  since such  operator is flavour-blind it does not contribute to any of these $\Delta F=1$ processes. It will be relevant though for the astrophysical constraints, see below.}
\begin{align}
	(c_{q_L})_{ij}\frac{\partial_{\mu} a}{f_a}\left( \bar{q}_{Li}\gamma^{\mu} q_{Lj}\right)
\end{align}
at low energies. Indeed, one obtains~\cite{Choi:2017gpf, MartinCamalich:2020dfe, Chala:2020wvs, Bauer:2020jbp}
\begin{align}
	16\pi^2 \frac{d {c_{q_L}}}{d\ln \mu}&=-\lambda_u c_{u_R}\lambda_u \Rightarrow \nonumber\\
	c_{q_L}&=\frac{\lambda_u c_{u_R}\lambda_u}{32\pi^2} \ln\left(\frac{f_a^2}{\mu^2}\right).
\end{align}
After EWSB, this operator leads to 
\begin{align}
	\frac{\partial_{\mu} a}{f_a} \left[(c_{u_L})_{ij}\bar{u}_{Li}\gamma^{\mu} u_{Lj}+(c_{d_L})_{ij}\bar{d}_{Li}\gamma^{\mu} d_{Lj}\right],
\end{align}
where $c_{u_L}=c_{q_L}$ and $c_{d_L}=V^{\dagger} c_{q_L}V$. More explicitly
\begin{align}
	(c_{d_L})_{ij}=\frac{1}{16\pi^2 v^2}V_{ri}^{\ast}(\mathcal{M}_u)_{rr} (c_{u_R})_{rs}  (\mathcal{M}_u)_{ss} V_{sj}\ln\left(\frac{f_a^2}{\mu^2}\right).
\end{align}
This WC is responsible for the exotic decays $B\to K a$, $B\to \pi a$ and $K\to \pi a$, where $B=B^{\pm,0},K=K^{\pm,0}$ and $\pi=\pi^{\pm,0}$.~\footnote{The form factors we use are computed in the isospin preserving limit so we do not make a difference between e.g. $B^{\pm} \to K^{\pm}a$ and $B^0\to K^0 a$.} The corresponding expressions can be read from equation~\eqref{eq:master} after taking $\mu\sim m_t$ and
\vspace{-0.3cm}
\begin{widetext}
	\begin{center}
	\begin{tabular}{llllrr}
		$(B \to K a)$,&$\varkappa_{MN}=(c_{u_{d_L}})_{32}$,& $m_{N}=m_{B}$,& $m_{M}=m_{K}$,& $f_0^{MN}(m_a^2)=f_0^{BK}(m_a^2)$ &\cite{Bailey:2015dka},\\
		$(B \to \pi a)$, & $\varkappa_{MN}=(c_{u_{d_L}})_{31}$,& $m_{N}=m_{B}$,& $m_{M}=m_{\pi}$,& $f_0^{MN}(m_a^2)=f_0^{B\pi}(m_a^2)$&\cite{Gubernari:2018wyi},\\
		$(K \to \pi a)$,& $\varkappa_{MN}=(c_{u_{d_L}})_{21}$,& $m_{N}=m_{K}$,& $m_{M}=m_{\pi}$,& $f_0^{MN}(m_a^2)=f_0^{K\pi}(m_a^2)$ &\cite{Carrasco:2016kpy}.
		\end{tabular}
	\end{center}
\end{widetext}

There are no constraints on the branching ratio $\mathrm{Br}(D\to\pi + \mathrm{invisible})$ to date. 
However, there are measurements  of the three-body meson decay $D^{+}\to (\tau^+\to \pi^+ \nu)\bar{\nu}$~\cite{Eisenstein:2008aa, Ablikim:2019rpl}. Since these analyses show the event distribution as a function of the missing mass squared $M_{\rm miss}^2$ (which would correspond to $m_a^2$ in the ALP case), one could recast them to constrain the branching ratio  $D^+\to \pi^+ a$.~\footnote{ See~\cite{Kamenik:2009kc} for the study of the long-distance lepton-mediated contributions to $B$ and $D$-meson semi-invisible decays.} This was done e.g. in~\cite{MartinCamalich:2020dfe} for the massless axion by concentrating on the bins with $M_{\rm miss}^2\le 0.05\,\rm{GeV}^2$. Since, as we will see, the total width of the ALP in all the benchmarks under consideration is very small, one can safely produce a similar bound for different values of $m_a$ by comparing the observed number of events with the predicted background for every bin having $M_{\rm miss}^2\ge 0$. More precisely, we derive 90\% CL on $\mathrm{BR}(D\to \pi a)$ by using the TLimit class of ROOT~\cite{Brun:1997pa}, which implements the CL$_s$ method~\cite{Read:2002hq} and  allows to include systematic errors in the background and signal. The bounds arising from~\cite{Eisenstein:2008aa} turn out to be stronger than those resulting from the use of the more recent experimental analysis in~\cite{Ablikim:2019rpl}.

Regarding exotic meson decays involving down quarks, there are several analysis focused on a massless axion $X^0$,  like $K^+\to \pi^+ X^0$~\cite{Adler:2008zza} and $B^{\pm}\to \pi^{\pm} X^0$, $B^{\pm}\to K^{\pm} X^0$~\cite{Ammar:2001gi}. There are no searches for a massive ALP $a$, with the exception of the recent analysis in~\cite{CortinaGil:2020fcx}, where bounds on $K^{+}\to \pi^{+}a$ as a function of $m_a$ were presented. Similarly to the $D^{\pm}\to \pi^{\pm} a$ case, we fill this gap by recasting existing searches on three-body decays, where the relevant kinematic information is provided. In particular, we derive constraints on $B\to Ka$ and $B\to \pi a$ by recasting the searches performed in~\cite{Lees:2013kla} and~\cite{Aubert:2004ws}, respectively. More specifically, we set 90\% CL on $B\to Ka$ by combining the observed number of events and the predicted background for $B^+\to K^+ \nu \bar{\nu}$ and $B^0\to K^0\nu \bar{\nu}$ for every $s_B=k^2/m_B^2=m_a^2/m_B^2$ bin in figure~5 of~\cite{Lees:2013kla} with the CL$_s$ method. Finally, for the case of $B\to \pi a$, we derive 90\% CL limits on $B\to \pi a$ with the CL$_s$ method  by comparing the observed number of events with the predicted background for every $p_{\pi}\equiv\sqrt{\vec{p}_{\pi}^{\, 2}}$ bin in the right panel of figure~4 in~\cite{Aubert:2004ws} (which is in one-to-one correspondence with the ALP mass via $m_a^2=m_B^2+m_{\pi}^2-2m_B\sqrt{m_{\pi}^2+\vec{p}_{\pi}^{\, 2}}$\,).

On the other hand, it is expected that Belle II will be sensitive to the SM $\mathrm{Br}(B^\pm \to K^\pm \nu\bar{\nu})=(4.0\pm0.5)\times 10^{-6}$ \cite{Buras:2014fpa} at 10\% accuracy with 50~ab$^{-1}$ of data \cite{Kou:2018nap}, whereas NA62 will measure the branching ratio $\mathrm{Br}(K^\pm\to\pi^\pm\nu\bar{\nu})$ to within 10\% of its SM value $\mathrm{Br}(K^\pm\to\pi^\pm\nu\bar{\nu})=(8.4\pm4.1)\times 10^{-11}$ \cite{Buras:2015qea,Martellotti:2015kna}. Regardless of whether these collaborations  publish limits directly on a two-body decay  or a recast of the three-body decay analysis is needed, such numbers represent a great improvement with respect to current bounds.~\footnote{See e.g.~\cite{Ertas:2020xcc} for a two-body interpretation of the NA62 prospects.}

\subsection{Radiative $J/\psi$ decays}
\begin{figure}[h]
	\begin{center}
		\includegraphics[width=0.48\linewidth]{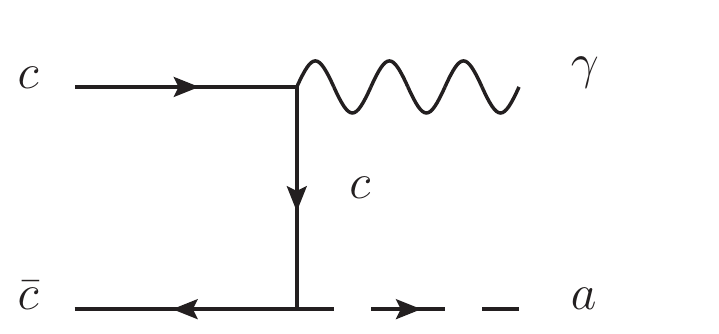}	
		\caption{Parton level diagram for the decay $J/\psi\to a\gamma$. }
		\label{fig:jpsi}
	\end{center}
\end{figure}
Diagonal ALP couplings to charm quarks are strongly constrained by charmonium decays like $J/\psi\to a\gamma$, as first proposed by   \cite{Wilczek:1977pj} and later studied by many others, see e.g.~\cite{Haber:1978jt, Haber:1987ua, Mangano:2007gi, Domingo:2008rr, Fayet:2008cn, Merlo:2019anv}. The parton-level diagram prompting such decay is shown in figure~\ref{fig:jpsi}. In order to absorb some of the QCD uncertainties of the calculation, it is convenient to normalize the corresponding branching ratio by the one of  $J/\psi\to \mu^+ \mu^-$, which is accurately measured $\mathrm{Br}(J/\psi \to \mu^-\mu^+)=5.973$~\% \cite{Ablikim:2013pqa}. One then obtains 
\begin{align}
	\frac{\mathrm{Br}(J/\psi\to a \gamma)}{\mathrm{Br}(J/\psi\to \mu^-\mu^+)}=&\frac{G_F m_c^2 v^2}{\sqrt{2}\pi\alpha_{\rm em}}\left(\frac{(c_{u_R})_{22}}{f_a}\right)^2\times\nonumber\\
	&\left(1-\frac{m_a^2}{m_{J/\psi}^2}\right) F,
\end{align} 
where  $F\sim \mathcal{O}(1/2)$ is a correction factor accounting for  QCD effects~\cite{Vysotsky:1980cz, Nason:1986tr}, contributions related to bound-state formation~\cite{Polchinski:1984ag, Pantaleone:1984ug} as well as relativistic corrections~\cite{Aznaurian:1986hi}. For the sake of concreteness, we will assume that $F=1/2$ henceforth. This leads to
\begin{align}
	\mathrm{Br}(J/\psi \to a \gamma) = (1.05\,\mathrm{GeV}^2)\,\left(\frac{(c_{u_R})_{22}}{f_a}\right)^2\left(1-\frac{m_a^2}{m_{J/\psi}^2}\right).
\end{align}
This decay has been searched for by the CLEO collaboration~\cite{Insler:2010jw}, which we will use to constrain the benchmark models at hand.
\section{Astrophysical and cosmological  bounds} 
\label{sec:cosm}
\subsection{Bounds from supernova SN1987a}
The observed neutrino burst due to the core-collapse supernova SN1987a can impose constraints on the ALP parameter space. Since neutrino emission constitutes the main cooling mechanism for the proto-neutron star resulting from the collapse, a too large ALP emission could compete with this cooling mechanism and eventually conflict the observed amount of neutrinos. Following~\cite{Raffelt:1996wa}, we will impose that the ALP luminosity in the proto-neutron star $L_a$ does not exceed the neutrino one $L_{\nu}$, i.e., $L_a\leq L_{\nu}=3\cdot 10^{52}\,\rm{erg}/s$.~\footnote{One should note, however, that the authors of ref.~\cite{Bar:2019ifz} have cast some serious doubts on supernova cooling bounds for ALPs.  }  

The ALP luminosity in the proto-neutron star  is given by~\cite{Chang:2016ntp, Chang:2018rso}
\begin{align}
	L_a=\int_{r\leq R_{\nu}} dV\,\int_{m_a}^{\infty} d\omega\, \left(\frac{dP_a}{dVd\omega}\right) \,e^{-\tau},
\end{align}
where $\omega$ is the ALP energy and $R_{\nu}\sim\mathcal{O}(40\,\rm{km})$ is the radius of the neutrinosphere, beyond which neutrinos free stream until arriving to the Earth, and we have taken into account the probability $e^{-\tau}$ for an ALP produced within the neutrinosphere to reach $R_{\rm far}\sim\mathcal{O}(100-1000\,\rm{km})$, after which neutrinos are not produced efficiently. If this is not the case, ALPs being produced within the neutrinosphere  get 'trapped' due to their large couplings and their energy is eventually converted back into neutrinos. Such probability is computed with the help of the optical depth~$\tau=\tau(m_a,\omega,r,R_{\rm far})$, for which we will take $R_{\rm far}=100\,\rm{km}$~\cite{Chang:2016ntp}.  In the above expression, $dP_a/dV d\omega$ is the ALP differential power. For the charming ALPs considered here, the main channel will be the bremsstrahlung process $N+N\to N+ N + a$, since the sole tree-level couplings are those to up-type quarks and therefore to nucleons. Such differential power is given by~\cite{Raffelt:1996wa, Raffelt:2006cw}
\begin{align}
	\frac{d P_a}{dV d\omega}=\frac{1}{2\pi^2} \omega^3 \Gamma_a e^{-\omega/T}\beta^2,
\end{align}
where $T$ is the temperature as a function of the radius, $\beta$ is a phase space factor $\beta=\sqrt{1-m_a^2/\omega^2}$ and $\Gamma_a$ is the ALP absorption width. The latter is given by~
\begin{align}
	\Gamma_a=\Gamma_a^{pp}+\Gamma_a^{nn}+\Gamma_a^{pn}+\Gamma_a^{np}
\end{align}
with~\cite{Chang:2018rso}
\begin{align}
	\Gamma_a^{N N^{\prime}}=\frac{c_{aNN}^2Y_NY_{N^{\prime}}}{4f_a^2}\frac{\omega}{2}\frac{n_B^2\sigma_{np\pi}}{\omega^2}\gamma_{\rm f}\gamma_{\rm p}\gamma_{\rm h},\quad N^{(\prime)}=n,p.
\end{align}
In the above equation, $c_{aNN}$ is the ALP-nucleon coupling, which reads (see appendix~\ref{app:cann} for more details)
\begin{align}
	c_{app}&=(c_{u_R})_{11}\left(0.75\pm0.03\right),\\
	c_{ann}&=(c_{u_R})_{11}\left(-0.51\pm0.03\right),
\end{align}
while $Y_{N^{(\prime)}}$ is the mass fraction of the nucleon $N^{(\prime)}$, $n_B$ is the baryon density, $n_B=\rho/m_N$, and $\sigma_{np\pi}$ is given by
\begin{align}
	\sigma_{np\pi}=4\alpha_{\pi}^2\sqrt{\pi T/m_N^5},
\end{align}
with $\alpha_{\pi}\approx 15$. For concreteness we take $Y_p=0.3$ and $Y_n=1-Y_p=0.7$. Moreover~\cite{Keil:1996ju},
\begin{align}
	1/\gamma_{\rm f}=1+\left(n_B\sigma_{np \pi}/(2\omega)\right)^2,
\end{align}
while we use $\gamma_{\rm p}=s(n_B,Y_{N},\omega/T, m_{\pi}/T)$ with $s$ given by eq.~(49) of~\cite{Hannestad:1997gc}.~\footnote{Note that at the end of the day, $s$ is divided by an extra factor $(1-\exp(-x))$ in order to preserve the detailed balance more explicitly.} On the other hand, following~\cite{Ertas:2020xcc}, we assume that~\cite{Bartl:2016iok}
\begin{align}
	\gamma_{\rm h}=-0.0726502\ln(\rho)+10^{10}/\rho^{ 0.9395710 }+2.5558616, 
\end{align}
where the density $\rho$ is expressed in $\rm{g}\, \rm{cm}^{-3}$. Similarly to~\cite{Ertas:2020xcc}, we assume for $\rho(r)$ and  $T(r)$ and the ''fiducial'' profiles of~\cite{Chang:2016ntp}
\begin{align}
	\rho(r)=\rho_c\times \left\{\begin{array}{ll}1+k_{\rho}(1-r/R_c)&r<R_c\\ (r/R_c)^{-\nu}&r\ge R_c\end{array}\right.,
\end{align}
\begin{align}
	T(r)=T_c\times \left\{\begin{array}{ll}1+k_{T}(1-r/R_c)&r<R_c\\ (r/R_c)^{-\nu/3}&r\ge R_c\end{array}\right.,
\end{align}
with $k_{\rho}=0.2$, $k_T=-0.5$, $\nu=5$, $R_c=10\,\rm{km}$, $T_c=30$\,MeV and $\rho_c=3\cdot 10^{14}\,\rm{g}/\rm{cm}^3$. We define $R_{\nu}$ as the distance at which the temperature is $3\,$MeV, obtaining $R_{\nu}=39.81\,\rm{km}$. Finally, for the optical depth we take~\cite{Ertas:2020xcc}
\begin{align}
	\tau=(R_{\rm far}-R_{\nu})\beta^{-1}\Gamma_a(R_{\nu})+\beta^{-1}\int_r^{R_{\nu}}d\tilde{r}\, \Gamma_a(\tilde{r}).
\end{align}
\subsection{Bounds from red giant burst}
The one-loop running of the dimension-5 effective Lagrangian will generate ALP couplings to electrons at low energy. Such couplings face very strong astrophysical bounds for small values of $m_a$. This effect can be particularly relevant when the ALP couples to the top quark, since it will contribute significantly to the running~\cite{Choi:2017gpf, MartinCamalich:2020dfe, Chala:2020wvs, Bauer:2020jbp}: 
\begin{align}
	16\pi^2\frac{d c_H}{d\ln \mu}&=-6\mathrm{Tr}\left(\lambda_u c_{u_R}\lambda_u\right)\Rightarrow \nonumber\\
	c_H&=\frac{3}{8\pi^2v^2}\mathrm{Tr}\left(\mathcal{M}_u c_{u_R}\mathcal{M}_u\right)\ln\left(\frac{f_a^2}{\mu^2}\right),
\end{align}
which leads after field redefinition to 
\begin{align}
	-c_H\frac{\partial_{\mu} a}{f_a}\left(\bar{l}_{L i}\gamma^{\mu}l_{L i}+2\bar{e}_{R i}\gamma^{\mu}e_{R i}\right).
\end{align}
After EWSB, these operators lead to 
\begin{align}
	\mathcal{L}\supset \frac{i c_H a}{f_a} m_{\ell} (\bar{\ell}\gamma_5 \ell) = i\, a \, g_{a\ell \ell} (\bar{\ell}\gamma_5 \ell),\quad \ell=e,\mu,\tau.
\end{align}
More explicitly, at $\mu\sim m_t$,
\begin{align}
	g_{aee}=\frac{c_H m_e}{f_a}=\frac{3 m_{e}}{8\pi v^2 f_a}\ln\left(\frac{f_a^2}{m_t^2}\right)
\sum_{i=1}^3 \left(\mathcal{M}_u\right)_{ii} (c_{u_R})_{ii}.\end{align}
This coupling is bounded by data from red giant bursts~\cite{Raffelt:2006cw, Feng:1997tn, DEramo:2018vss,Capozzi:2020cbu},  $g_{aee}\lesssim 1.6\cdot 10^{-13}$, for ALP masses below the temperature of the red giant.

\subsection{ALP lifetime, branching ratios and cosmological bounds}
Cosmological bounds are very sensitive  to the total decay width and the different branching ratios of the ALP, that we discuss in the following.  We do this both for small values of the ALP mass, where QCD is confined and one can use chiral perturbation theory, as well as for larger values  where the dominant ALP decays can be computed using quark-hadron duality~\cite{Poggio:1975af, Shifman:2000jv}. Following~\cite{Renner:2018fhh}, we determine the energy scale separating both pictures by demanding that the total decay width to hadrons or SM quarks is of the same order in both regimes, which leads to $\sim 1$\,GeV for the benchmark models at hand.

For small masses, $m_a\lesssim 1\,$GeV, the ALP decays to two photons via the mixing \eqref{eq:mmix} and the subsequent decay $\pi\to \gamma\gamma$, plus one-loop contributions coming from the integration of heavy quarks. This leads to~\cite{Bauer:2017ris, Aloni:2018vki}
\begin{align}
	\Gamma(a\to \gamma \gamma)&\approx \frac{\alpha_{\rm em}^2 m_a^3}{(4\pi)^3 f_a^2}\left| \sum_{i=2}^3 \frac{4}{3} (c_{u_R})_{ii} B_1(\tau_{i})\right.\nonumber\\
	&\left.-\frac{m_a^2}{2(m_{\pi}^2-m_a^2)}(c_{u_R})_{11}\right|^2,
\end{align}
where $B_1(\tau_i)=1-\tau_i f^2(\tau)$, $\tau_i=4 (\mathcal{M}_u)_{ii}^2/m_a^2$, and 
\begin{equation}
	f(\tau_i)=\left\{\begin{array}{cc}\arcsin(1/\sqrt{\tau_i})& \tau_i\ge 1\\ \frac{\pi}{2}+\frac{i}{2}\ln\left(\frac{1+\sqrt{1-\tau_i}}{1-\sqrt{1-\tau_i}}\right)&\tau_i<1 \end{array}.\right. 
\end{equation}
This decay channel will be the only one present, whenever the one-loop lepton decays $a\to \ell^+ \ell^-$  are kinematically closed. The leptonic decay widths can be written as
\begin{align}
	\Gamma(a\to \ell^+\ell^-)=\frac{m_a m_{\ell}^2}{8\pi f_a^2}\sqrt{1-\frac{4m_{\ell}^2}{m_a^2}}|c_H|^2.
\end{align}
The diphoton final state will dominate over the $e^{+}e^{-}$ and $\mu^+\mu^-$ decays in models where $(c_{u_R})_{33}$ is absent or negligible. Otherwise, once $a\to e^+e^-$ and $a\to \mu^+\mu^-$ open up, they will become the main ALP decay channel, at least for large values of $f_a$ leading  to log-enhanced  $g_{a\ell\ell}$ couplings. At any rate, the $m_a^3$ dependence of the diphoton decay will make $\Gamma(a\to \gamma\gamma)$ increase faster than the dilepton decay width with increasing values of $m_a$. In some cases, this can turn such decay channel into the leading one, once more, for larger ALP masses  before $a\to 3\pi$ opens kinematically.  Such decay channel will always dominate the $a\to \gamma\gamma$ final state in our models, with its decay width reading~\cite{Bauer:2017ris}
\begin{align}
	\Gamma(a\to\pi^a\pi^b\pi^0) &=\frac{\pi}{12}\frac{m_a m_\pi^4}{f_a^2 f_\pi^2}\left[\frac{(c_{u_R})_{11}}{32\pi^2}\right]^2g_{ab}\left(\frac{m_\pi^2}{m_a^2}\right),
\end{align}
with 
\begin{align}
g_{00}(x)&=\frac{2}{(1-x)^2}\int_{4x}^{(1-\sqrt{x})^2}dz\sqrt{1-\frac{4x}{z}}\lambda^{1/2}(1,z,x),\\
	g_{+-}(x)&=\frac{12}{(1-x)^2}\int_{4x}^{(1-\sqrt{x})^2}dz\sqrt{1-\frac{4x}{z}}(z-x)^2\nonumber\\
	&\times \lambda^{1/2}(1,z,x),
\end{align}
and $\lambda(a,b,c)=a^2+b^2+c^2-2(ab+ac+bc)$. However, when $(c_{u_R})_{33}$ is sizeable and/or $f_a$ high enough to have significant $g_{a\ell\ell}$ couplings, $a\to \mu^+\mu^-$ can be the dominant decay channel in this region of ALP masses.

For masses $m_a\gtrsim 1\,$GeV, equation~\eqref{lag:alp} becomes invalid  and the dominant ALP decays into hadrons can be computed using quark-hadron duality. The ALP will then decay into gluons and quarks. While the first decay is loop induced, the tree-level decay into quarks will be suppressed by the light Yukawa couplings. Close to threshold, $a\to \bar{q}^{(\prime)}q$ will be the leading decay channel,  whereas for larger values of $m_a$ $a\to gg$ will take over. In this regime, the $a\to \gamma\gamma$ decay will always be sub-leading and read
\begin{align}
	\Gamma(a\to \gamma \gamma)&\approx \frac{\alpha_{\rm em}^2 m_a^3}{(4\pi)^3 f_a^2}\left| \sum_{i=1}^3 \frac{4}{3} (c_{u_R})_{ii} B_1(\tau_{i})\right|^2.
\end{align}

 \begin{figure*}[t!]
 	\begin{center}
 		\includegraphics[width=0.47\linewidth]{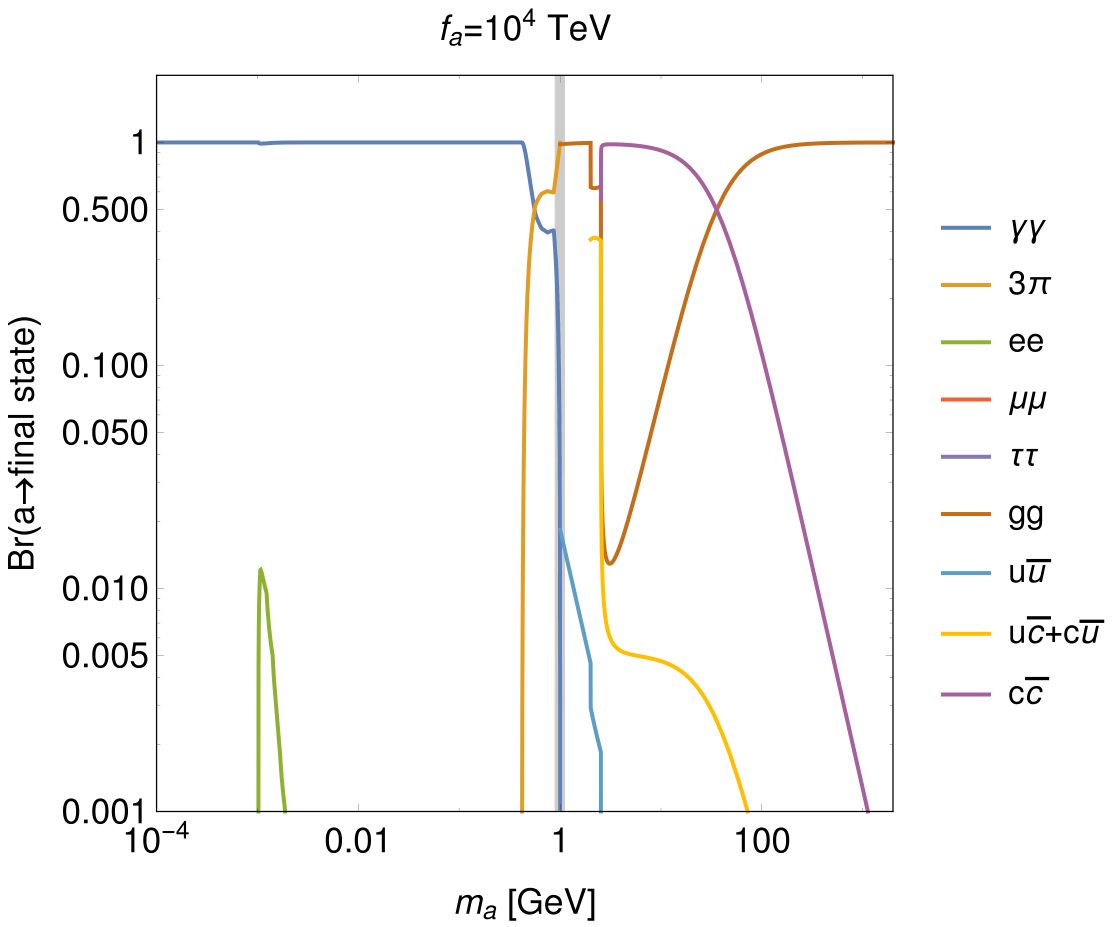}
 		\quad
 		\includegraphics[width=0.47\linewidth]{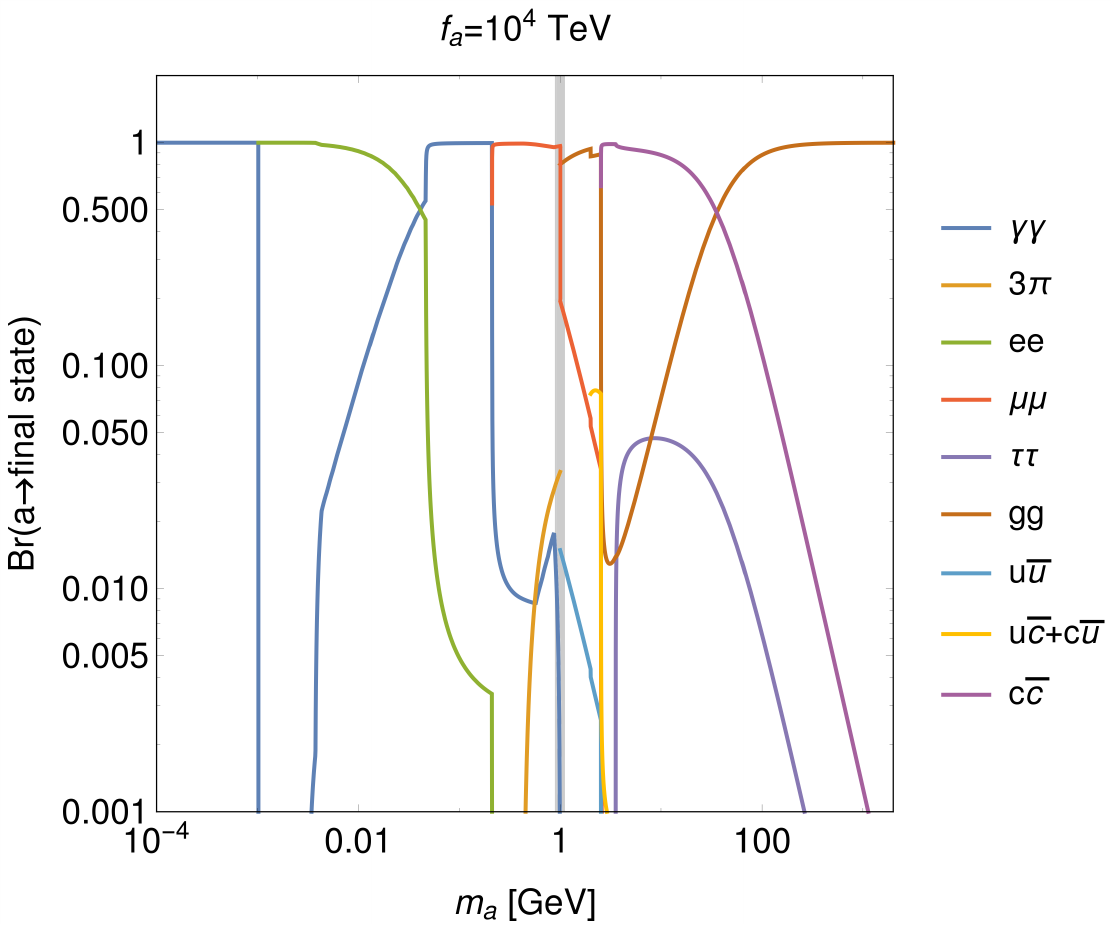}
 		\quad\includegraphics[width=0.47\linewidth]{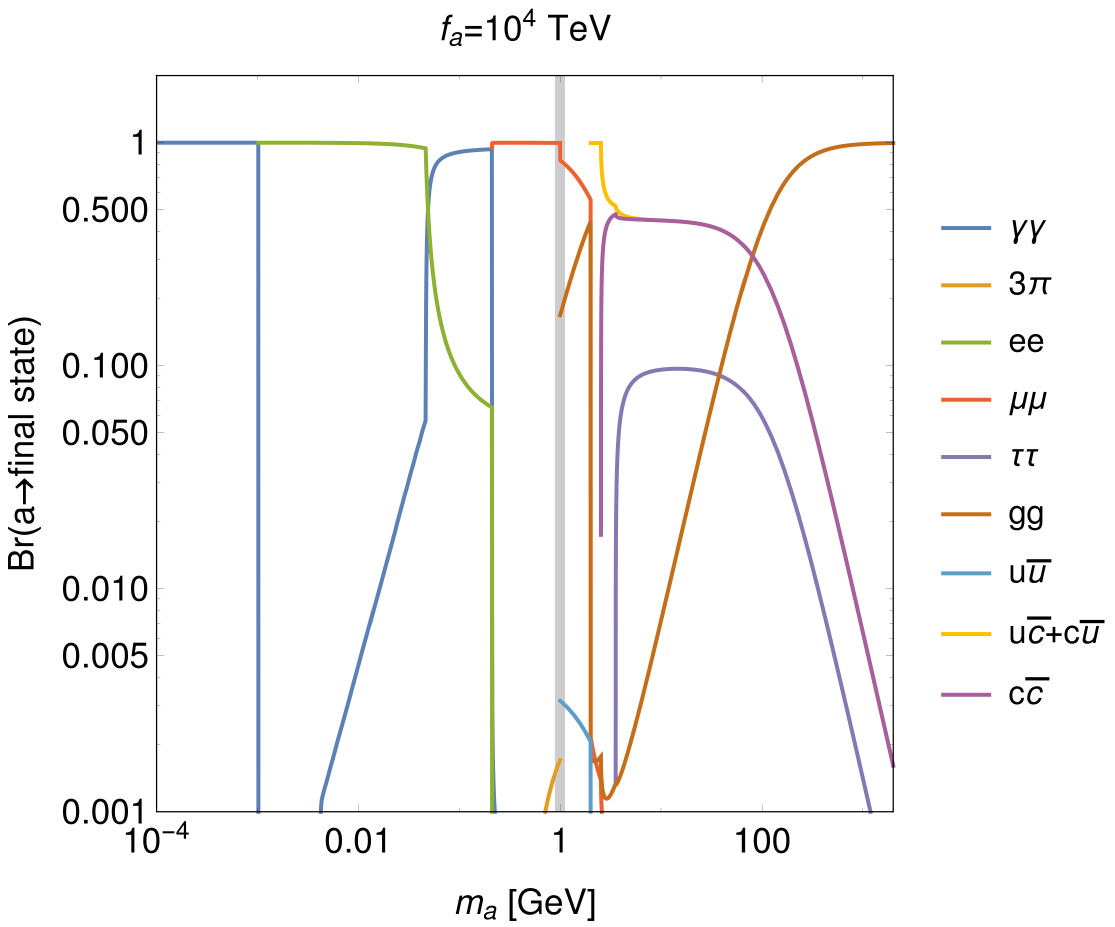}
 		\quad
 		\includegraphics[width=0.47\linewidth]{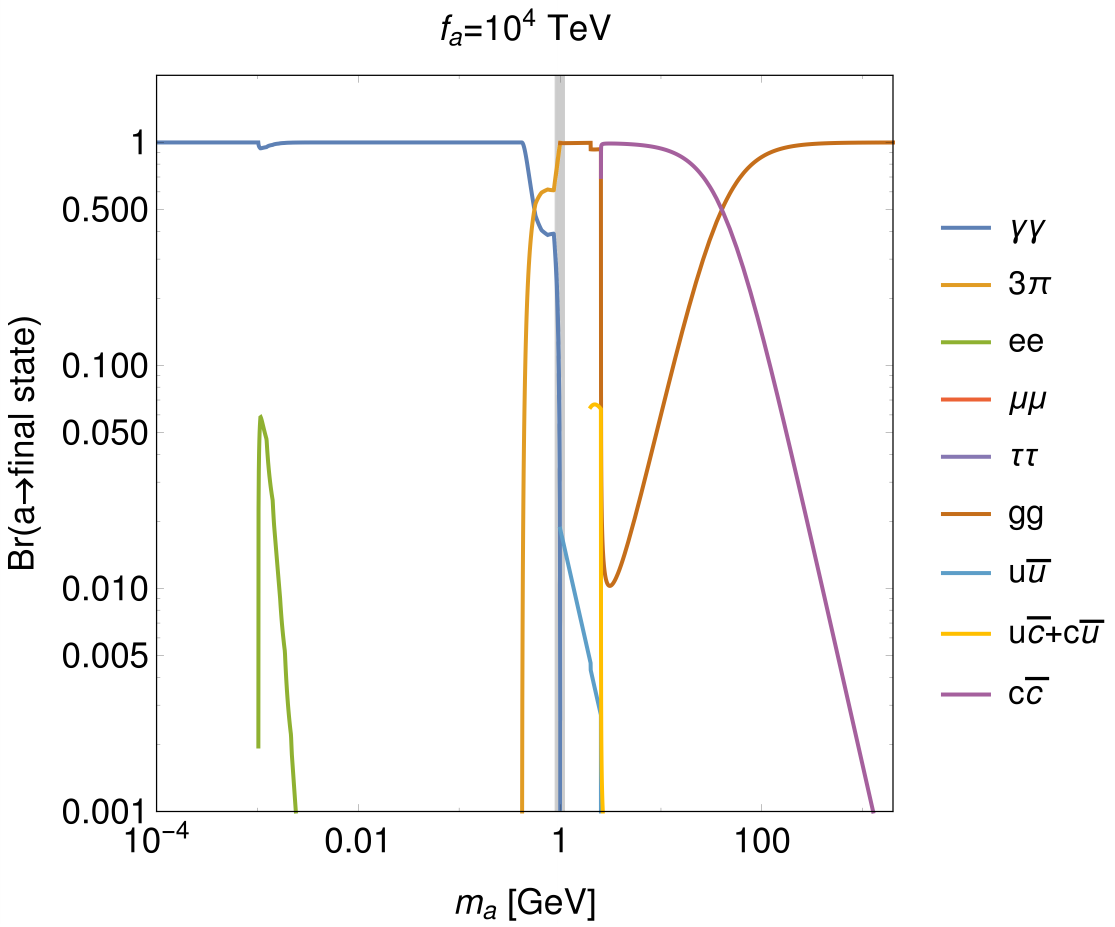}	
		\caption{Branching ratios of the ALP as a function of its mass, $m_a$, for the different benchmark models. The top panels correspond to the dark-QCD inspired case with $a=\pi_{D_3}$ (top left) and $a=\pi_{D_8}$ (top right), whereas the bottom panels show the anarchic scenario (bottom left) and the FN motivated benchmark (bottom right), respectively. In all cases, we have assumed $f_a=10^4$~TeV. Moreover, for the dark-QCD motivated scenarios illustrated in the top panels we have taken $\kappa_0=1$. We illustrate with a gray narrow band around 1~GeV, the matching between the calculations performed using chiral perturbation theory  and with quark-hadron duality. }
 		\label{fig:brs}
 	\end{center}
 \end{figure*}

We show in figure~\ref{fig:brs} the different branching ratios for the models at hand for $f_a=10^4$~TeV, assuming that $\kappa_0=1$ in the dark-QCD motivated benchmarks. We can see that in the cases where the ALP coupling to top quarks is absent or suppressed (top-left and bottom-right plots), $a\to \gamma\gamma$ is the leading decay mode for most of the region $m_a\lesssim 1$~GeV until $a\to 3\pi$ is kinematically allowed. In the cases where this coupling is present (top-right and bottom-left plots), and for this choice of $f_a$,  $a\to e^+e^-$ becomes the leading channel until $a\to \mu^+\mu^-$ opens up. For ALP masses $\gtrsim 1$~GeV, decays into hadrons are by far the dominant channels, with $a\to gg$ leading at large masses. 

Now that we have computed the different branching ratios and lifetimes, we can evaluate the impact of the cosmological constraints. It is important to stress that most of them are derived assuming that $a$ only interacts with photons. However, as discussed in~\cite{Cadamuro:2011fd}, one can apply these cosmological bounds to the more general case where other couplings are present.  At the end of the day, we will be able to recast the limits from~\cite{Cadamuro:2011fd, Millea:2015qra,Depta:2020wmr} by using the ALP lifetime  $1/\Gamma$, with $\Gamma$ the  ALP total decay width. These bounds include the possible impact on $N_{\rm eff}$, potential  distortions of the cosmic microwave-background spectrum as well as modifications of the predicted big-bang nucleosynthesis, see~\cite{Cadamuro:2011fd, Millea:2015qra,Depta:2020wmr}.

 	\begin{table*}[t]
		\begin{center}
 		\begin{tabular*}{0.8\textwidth}{@{\extracolsep{\fill}}lcccc}
 			\colrule\colrule
 			\textbf{Experiment}&distance from IP&length of decay volume &radius/opening angle& $N_D$	\\\toprule
 			FASER&480~m&1.5~m&0.1~m&$1.1\times10^{15}$ \\\colrule
 			FASER2&480~m&5~m&1~m&$2.2\times10^{16}$ \\\colrule
 			MATHUSLA&68~m downstream,&100~m&25~m high&$2.2\times10^{16}$\\
 			& 60~m above&&&\\ \colrule
 			NA62&80~m&65~m&$\theta_{\rm max}=0.05$&$2\times10^{15}$\\\colrule
 			SHiP&60~m&50~m&2.5~m&$6.8\times10^{17}$\\\colrule
 			CHARM&480~m&35~m&$0.0068< \theta<0.0126$&$4.08\times10^{15}$\\ 
 			\colrule\colrule
 		\end{tabular*}
 		\caption{Detector parameters for the different fixed-target experiments and LHC forward detectors considered.}
 		\label{tab:detectors}
		\end{center}
 	\end{table*}	

Indeed, the bounds can directly be applied when the decay to lepton pairs is dominant, i.e. when there is a sizeable coupling of the ALP with the top quark, provided one interprets $1/\Gamma$ as the total lifetime.~\footnote{When $a\to\mu^+\mu^-$ dominates, this slightly over-estimates the excluded region, since the subsequent decay of the muon also heats the neutrino bath, which reduces the impact on $N_{\rm eff}$. }
In the cases where  $a\to 3\pi$ dominates, bounds from $^4$He overproduction -- the dominant constraint in this region -- will still hold regardless of the changes in the branching ratios, since only a minimal amount of charged pions is enough for this bound to apply. For even larger masses, ALP decays into hadrons will eventually make its lifetime shorter than a second, making nucleosynthesis constraints harmless.    Therefore, even for these masses, we can apply the corresponding bounds if we interpret $\tau$ as the total lifetime.

 \section{Collider probes}
 \label{sec:coll}

 \subsection{Fixed target experiments}
 The main production mode for charming ALPs  
 at fixed-target experiments is the decay of $D$ mesons. We consider NA62 \cite{NA62:2017rwk} and the proposed SHiP experiment \cite{Alekhin:2015byh} as possible detection experiments. We also consider the bounds imposed by the CHARM experiment in \cite{Bergsma:1985qz}. The geometrical outlines of the experiments are listed in table~\ref{tab:detectors}. 

 NA62 operating in beam dump mode, meaning the target is lifted so that the 400~GeV proton beam hits the Cu collimator located 20\,m downstream, can be used to search for hidden sector particles \cite{Dobrich:2017yoq}.  A short run in beam dump mode in November 2016 provided useful information about the backgrounds. It was found that an upstream veto in front of the decay volume could reduce the background to nearly zero \cite{Lanfranchi:2017wzl}.
 The layout of the SHiP detector is proposed with the aim of reducing the beam-induced backgrounds to 0.1 events \cite{Ahdida:2654870,Alekhin:2015byh}, so that 3 decay events correspond to the expected exclusion region at over 95\% CL.
 
 The total number of dark pions decaying inside the respective decay volume is
 \begin{align}
	 N_{a}=N_D\cdot \mathrm{Br}(D\to\pi a)\cdot\varepsilon_{\rm geom}\cdot F_{\rm decay}\,,
 \label{eq:ND}
 \end{align}
 with  $\varepsilon_{\rm geom}$ the geometric acceptance, defined as the fraction of ALPs with lab frame momentum at the acceptance angle of the respective detector and $F_{\rm decay}$ the fraction of ALPs that decay inside the decay volume of the respective detector. $F_{\rm decay}$ and $\varepsilon_{\rm geom}$ are calculated following \citep{Renner:2018fhh}. The $D$ meson momentum distribution for  SHiP is taken from \cite{CERN-SHiP-NOTE-2015-009}. The same distribution is used for the NA62 case as the proton beam is the same.
 
The CHARM experiment searched for ALPs decaying into pairs of photons, electrons and muons. In \cite{Bergsma:1985qz} no events where found, so that we set a bound at 90\% CL at $N_{\rm obs}=2.3$ events \cite{Clarke:2013aya}. We assume that for our model the main production channel for ALPs is the decay $D^\pm\to\pi^\pm a$. CHARM also has a 400~GeV proton beam, so we again use the same momentum distribution for the $D$ mesons. The number of protons on target is $2.4\times10^{18}$ \cite{Bergsma:1985qz} and such a proton beam has a probability of $1.7\times 10^{-3}$ to produce a pair of $c$ quarks \cite{Abt:2007zg}, leading to $4.08\times 10^{15}$ produced $D$ mesons. Multiplying eq.~\eqref{eq:ND} by $\sum_{i=\gamma,e,\mu}Br(a\to ii)$ to take into account that in~\cite{Bergsma:1985qz} only the $\gamma\gamma$, $ee$ and $\mu\mu$ final states were searched for we use the same procedure as for NA62 and SHiP to impose the 90\% CL bound from CHARM.

 \subsection{LHC forward detectors}

 FASER \cite{Ariga:2018uku} and the proposed MATHUSLA detector \cite{Alpigiani:2018fgd,Alpigiani:2020tva} are designed to detect long-lived particles produced in proton-proton collisions at LHC. Hadron collisions have the advantage of additional production modes, such as production via gluon fusion. The detector parameters are given in table~\ref{tab:detectors}. We focus again on the production via $D$ meson decays.  The meson momentum distribution was simulated using  FONLL with CTEQ6.6 \cite{Cacciari:1998it}. As for SHiP and NA62 the number of dark pions decaying inside the decay volume can be calculated using eq.~\eqref{eq:ND}. For LHC run-3 $N_D =1.1\times10^{15}$, which increases by a factor 20 at the High Luminosity (HL)-LHC. FASER will operate at LHC run-3, while FASER2 and MATHUSLA are under consideration for the HL-LHC. Following the procedure as described for SHiP and NA62 the number of dark pions decaying inside FASERs, FASER2s and MATHUSLAs decay volume is calculated. At least three events must decay inside the respective decay volume for a discovery.

 \section{Results}
 \label{sec:res}

 \begin{figure*}[t]
 	\begin{center}
 		\includegraphics[width=0.47\linewidth]{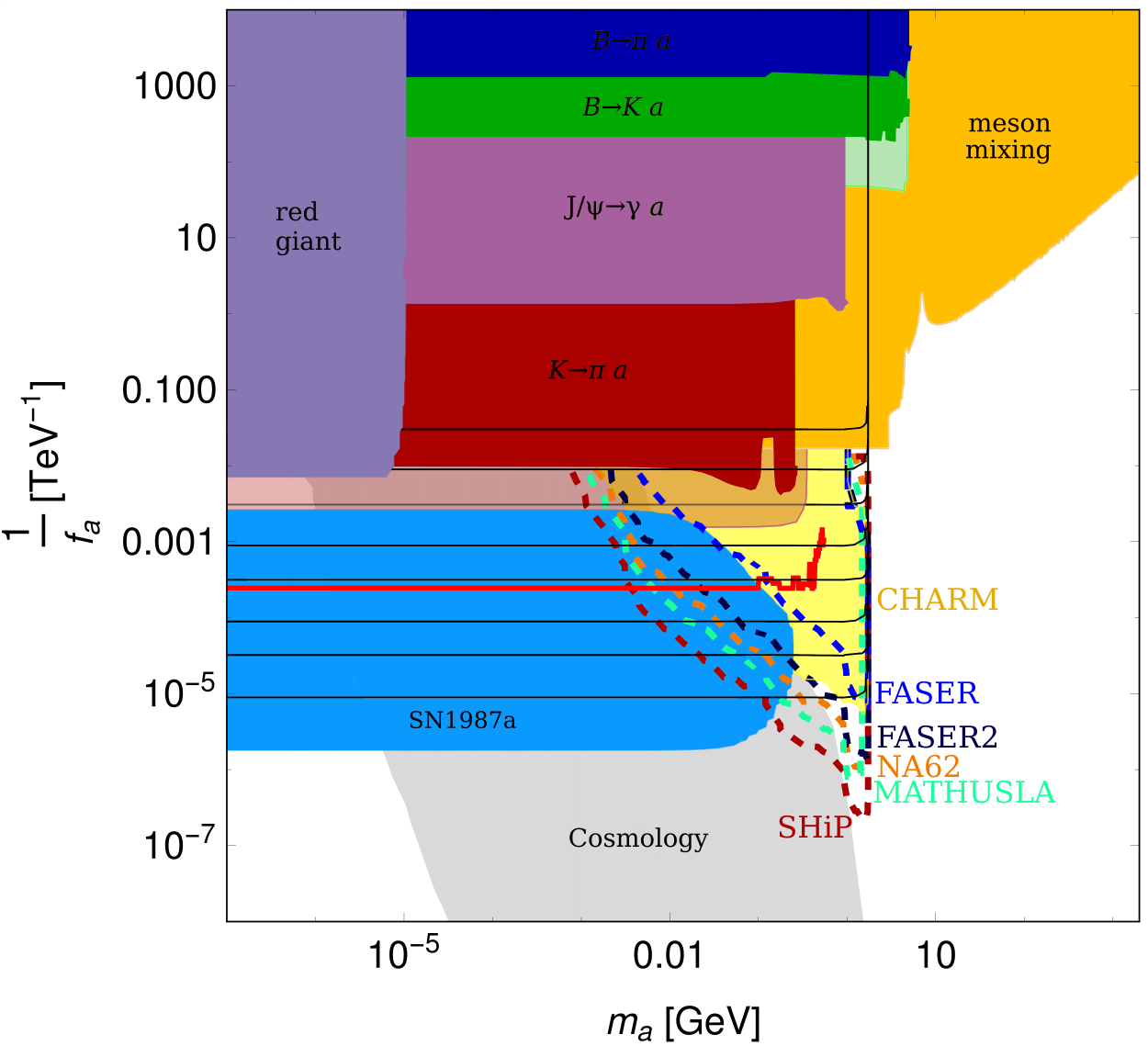}
 		\quad
 		\includegraphics[width=0.47\linewidth]{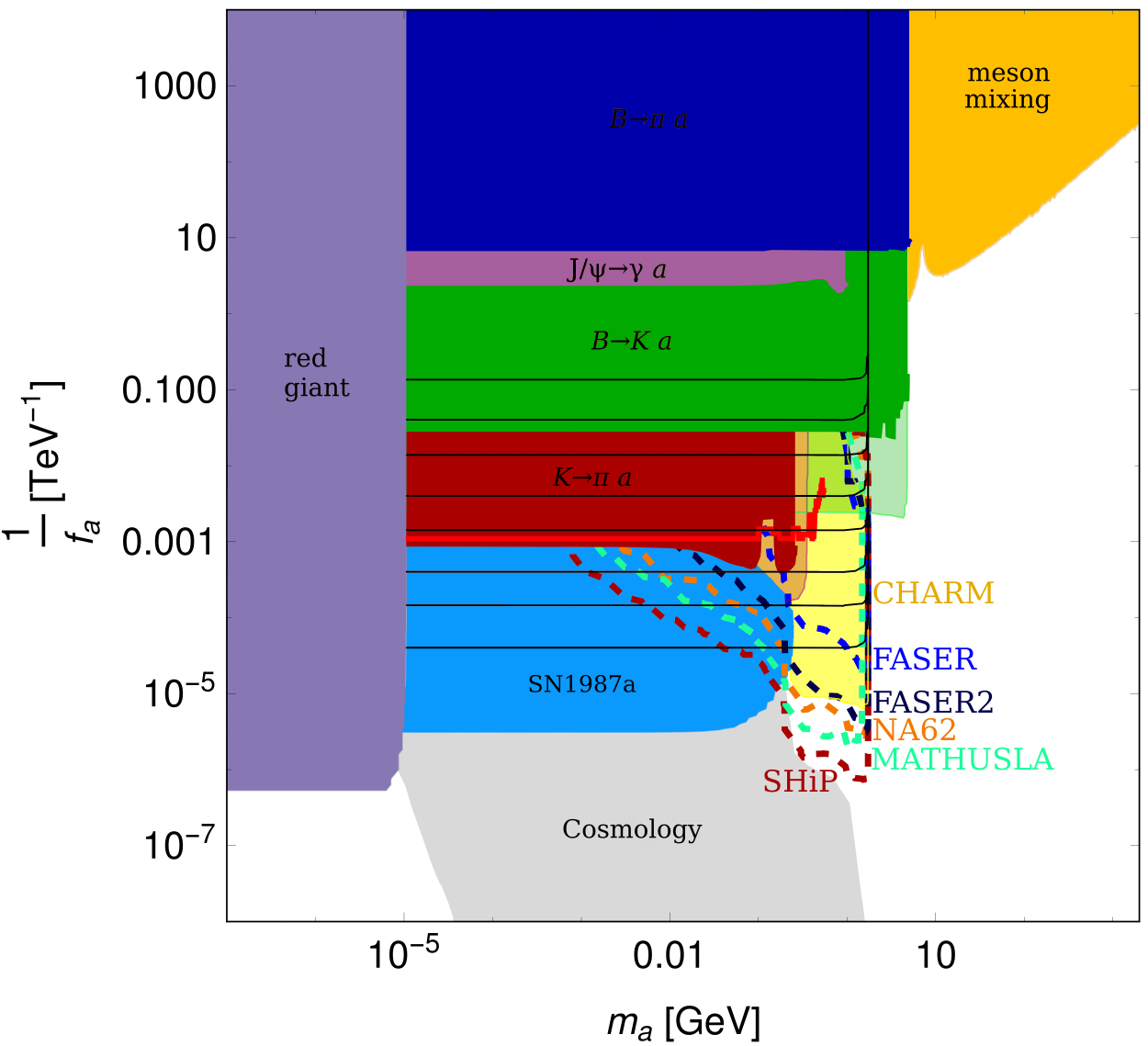}
 		\quad\includegraphics[width=0.47\linewidth]{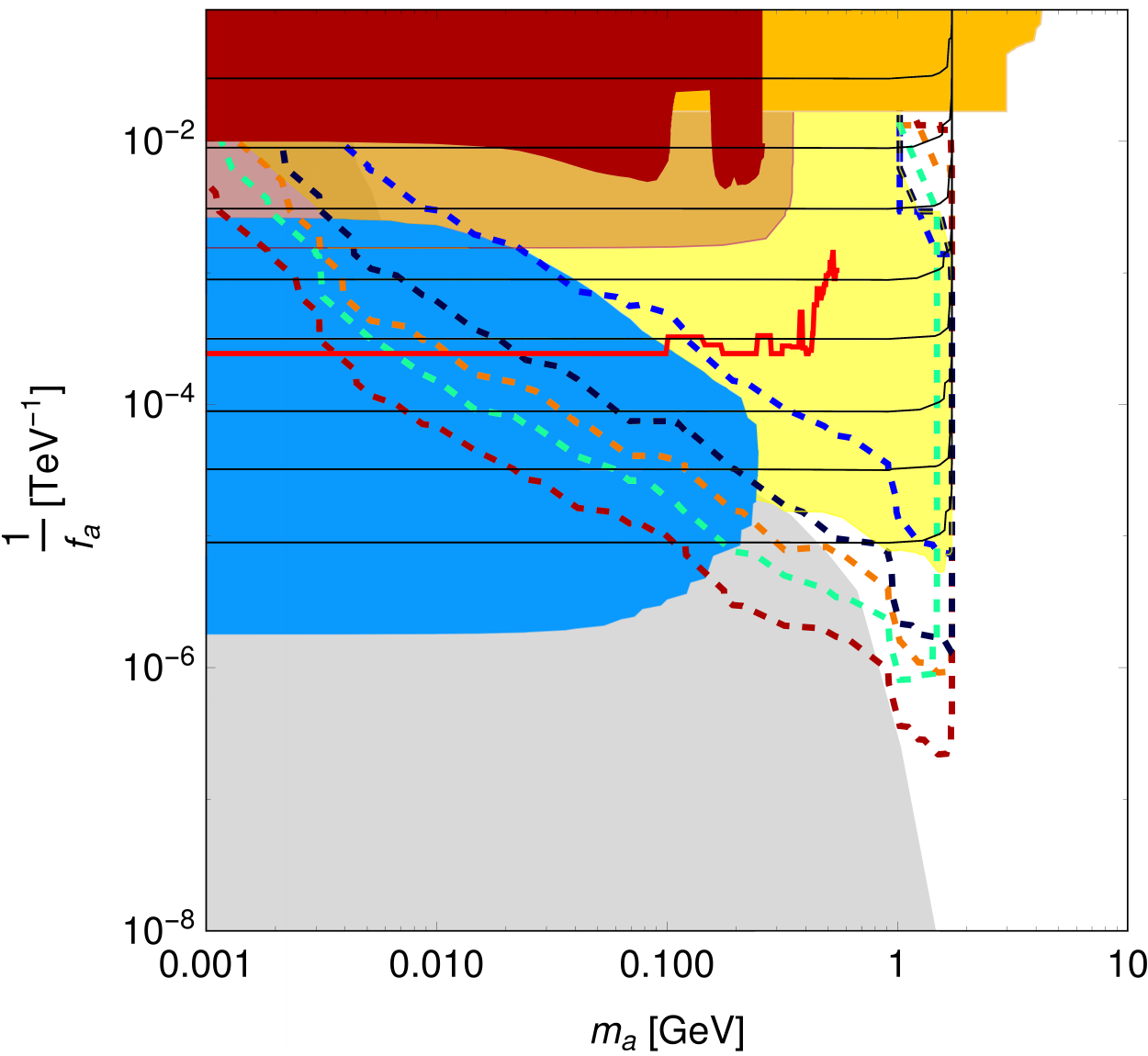}
 		\quad
 		\includegraphics[width=0.47\linewidth]{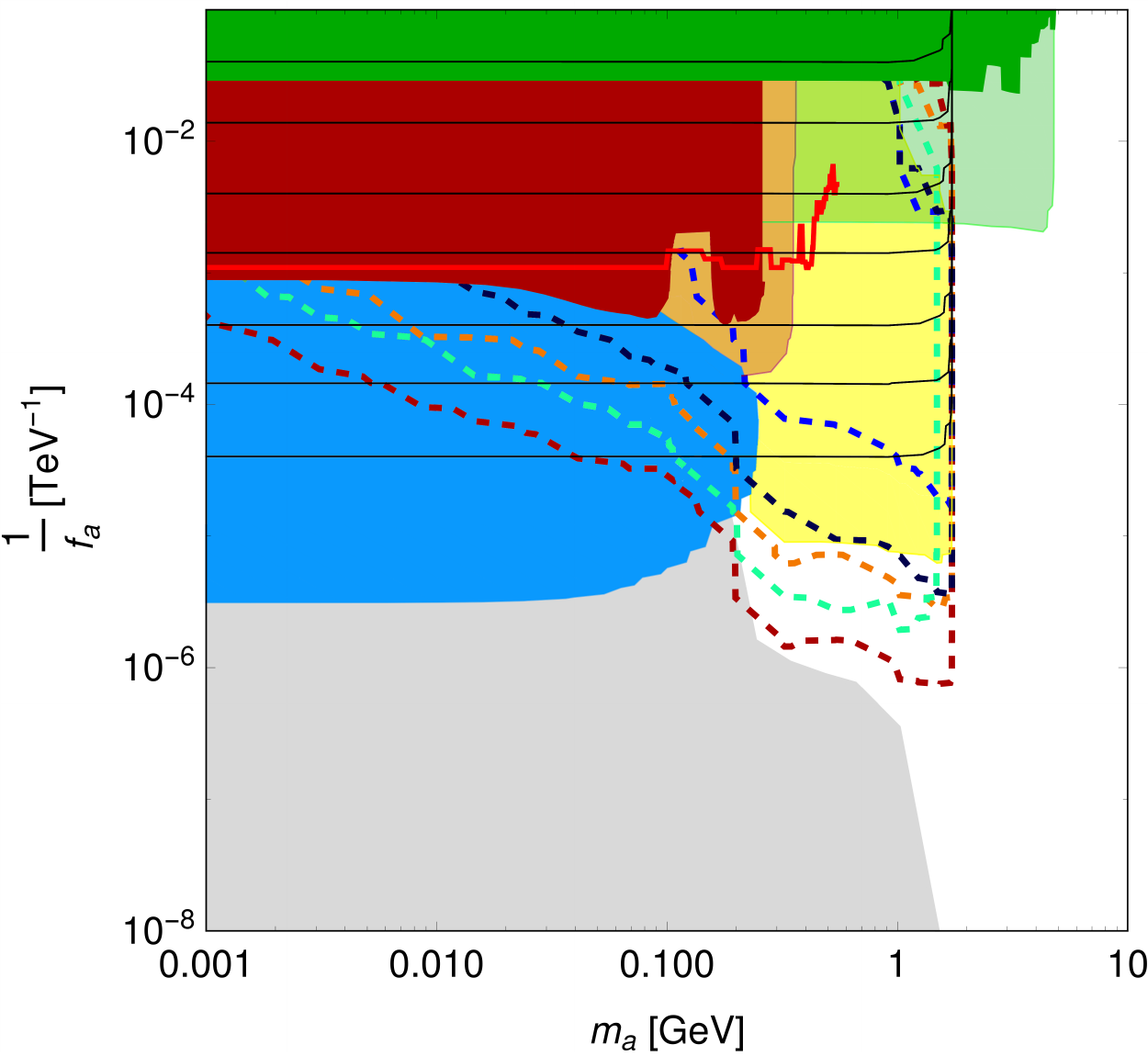}	
		\caption{Experimental constraints and expected bounds on $1/f_a$ as a function of $m_a$ for the dark-QCD inspired models. Left panels correspond to the case $a=\pi_D^3$, while right panels illustrate the scenario $a=\pi_D^8$. Light green and light red areas correspond to the expected  constraints coming from Belle II and NA62, respectively. We show by a red line the constraints arising from the recast of the three-body decay $D^+\to (\tau^+\to \pi^+ \bar{\nu})\nu$, whereas the impact of a direct measurement of $\mathrm{Br}(D\to \pi a)$ is represented by black lines, with values going from $10^{-1}$ to $10^{-8}$, each one a decade smaller. Dashed lines correspond to different fix-target experiments and collider probes of the model, see the main text for more details. Lower panels zoom in the regions where upcoming experiments are sensitive. In order to evaluate the logarithms coming from the one-loop running we further assume $\kappa_0=1$.}
 		\label{fig:ALP_results}
 	\end{center}
 \end{figure*}

\begin{figure*}[t]
	\begin{center}
		\includegraphics[width=0.47\linewidth]{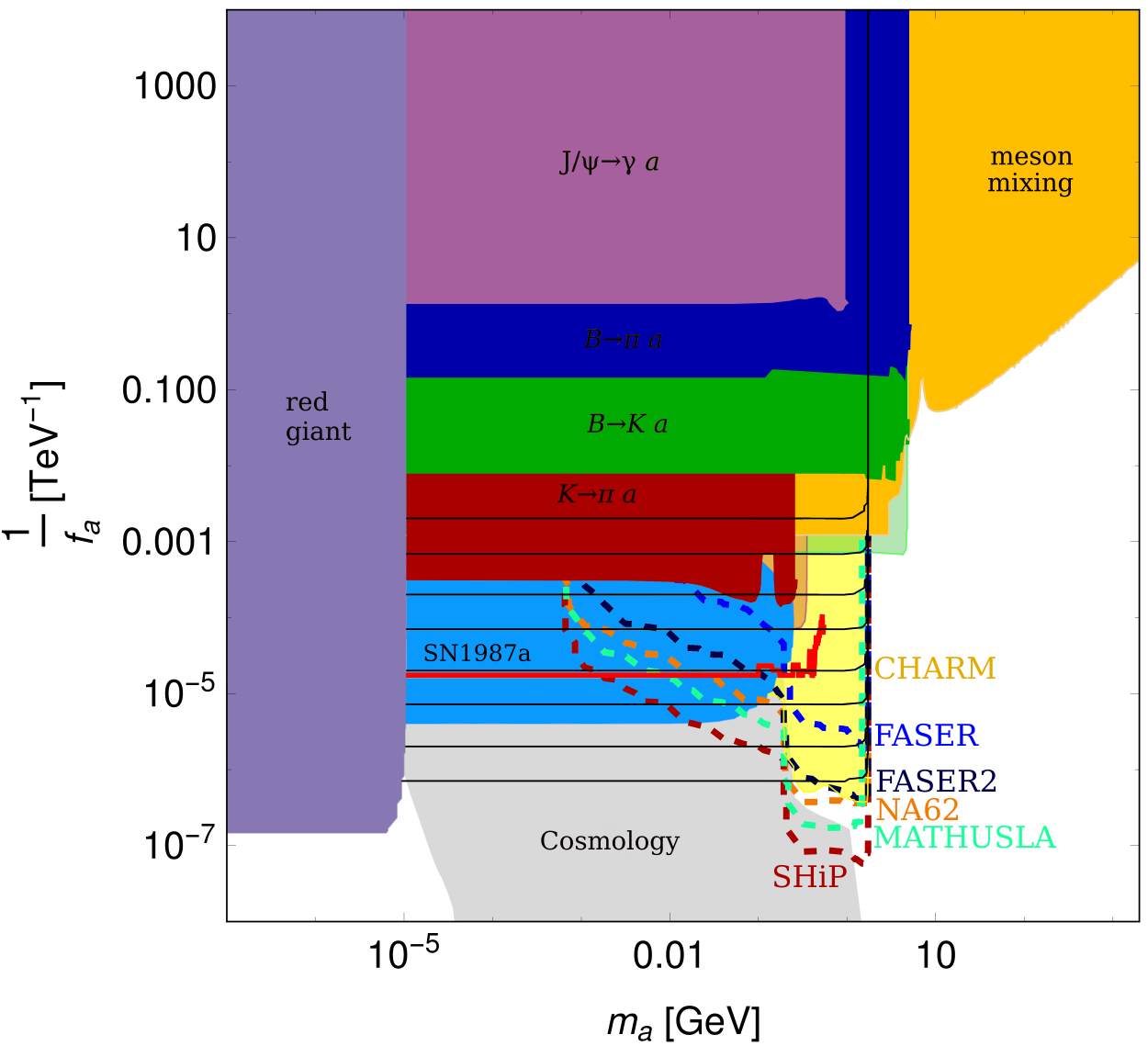}
		\quad
		\includegraphics[width=0.47\linewidth]{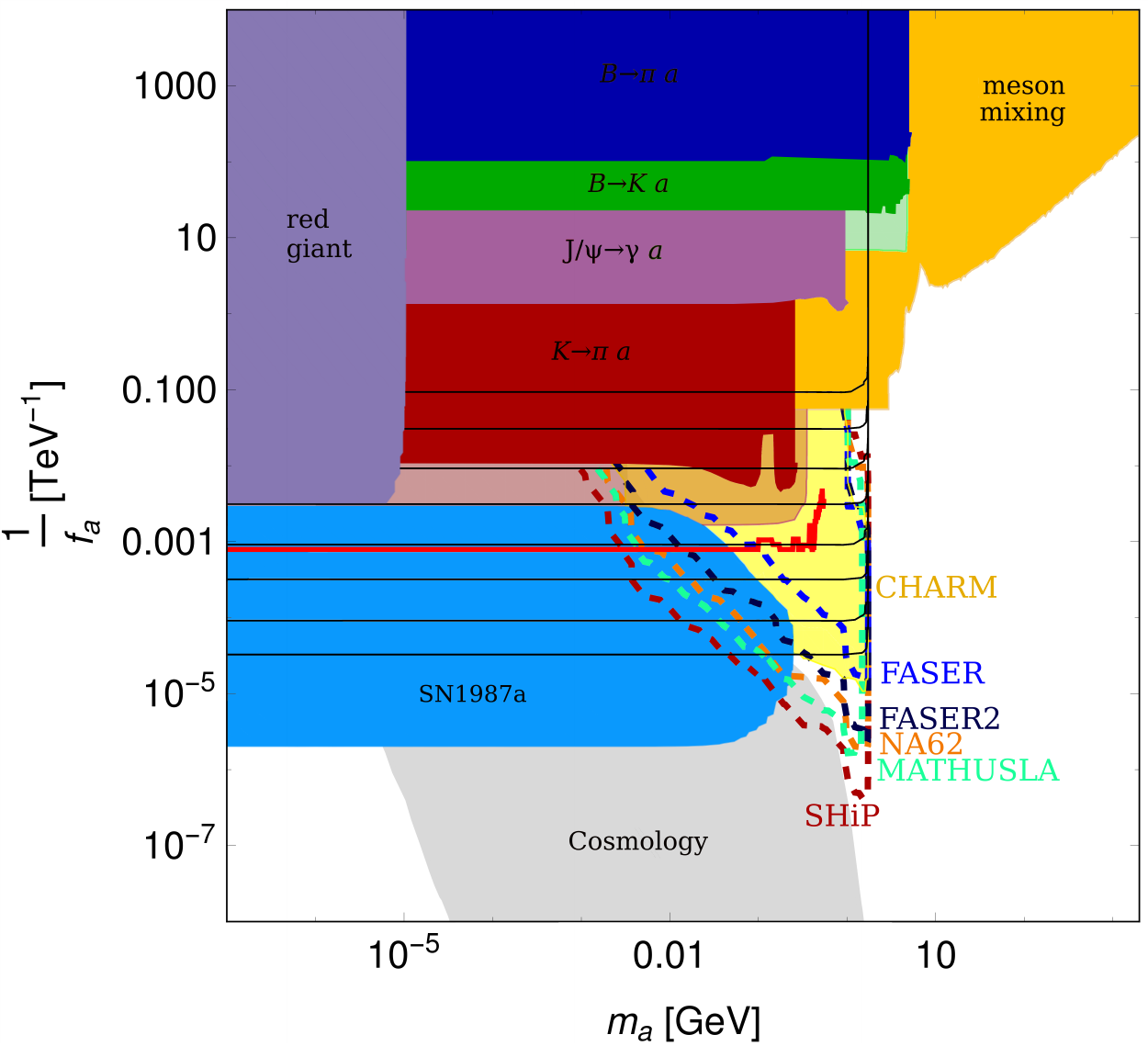}	
		\quad
		\includegraphics[width=0.47\linewidth]{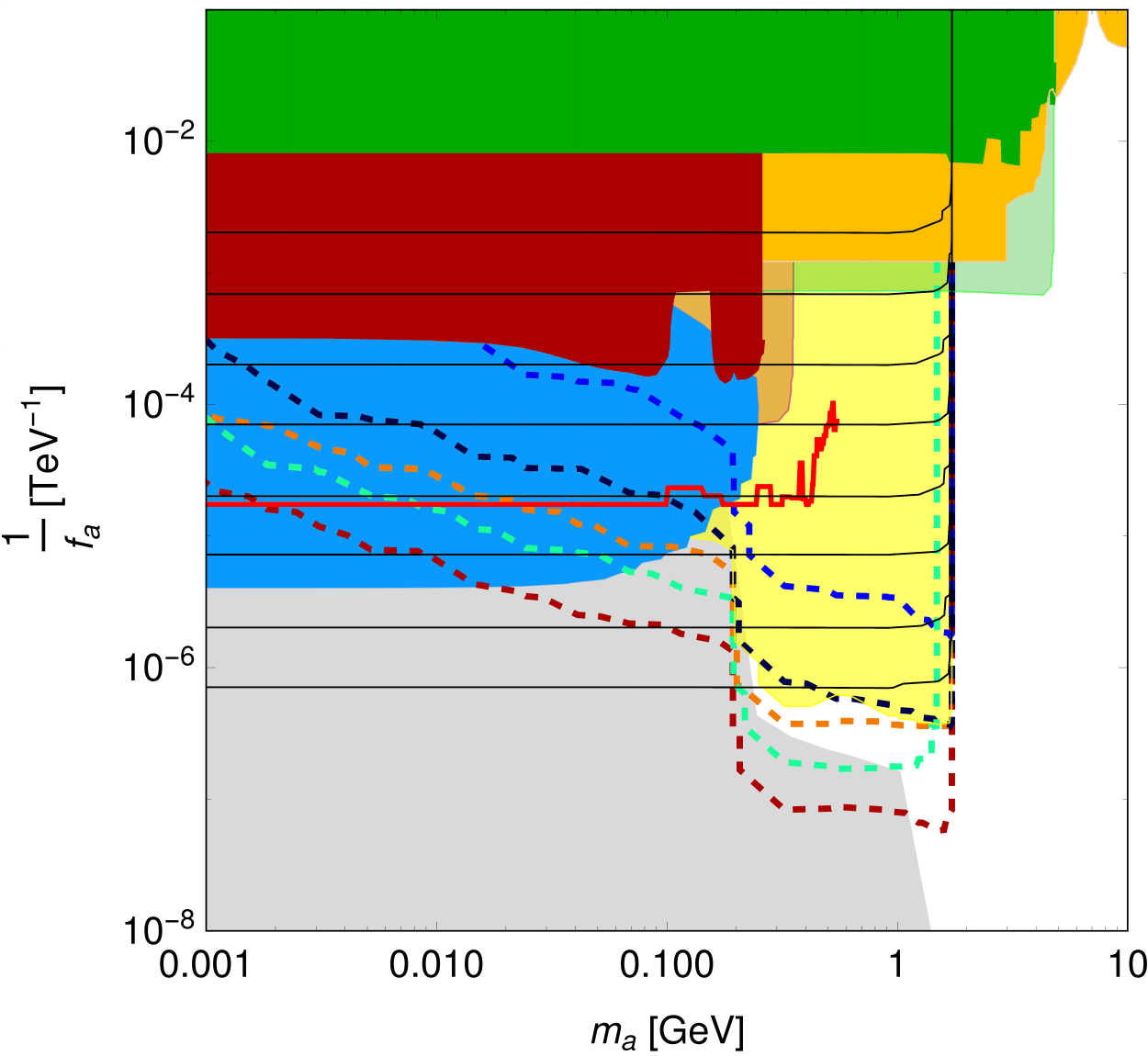}
		\quad
		\includegraphics[width=0.47\linewidth]{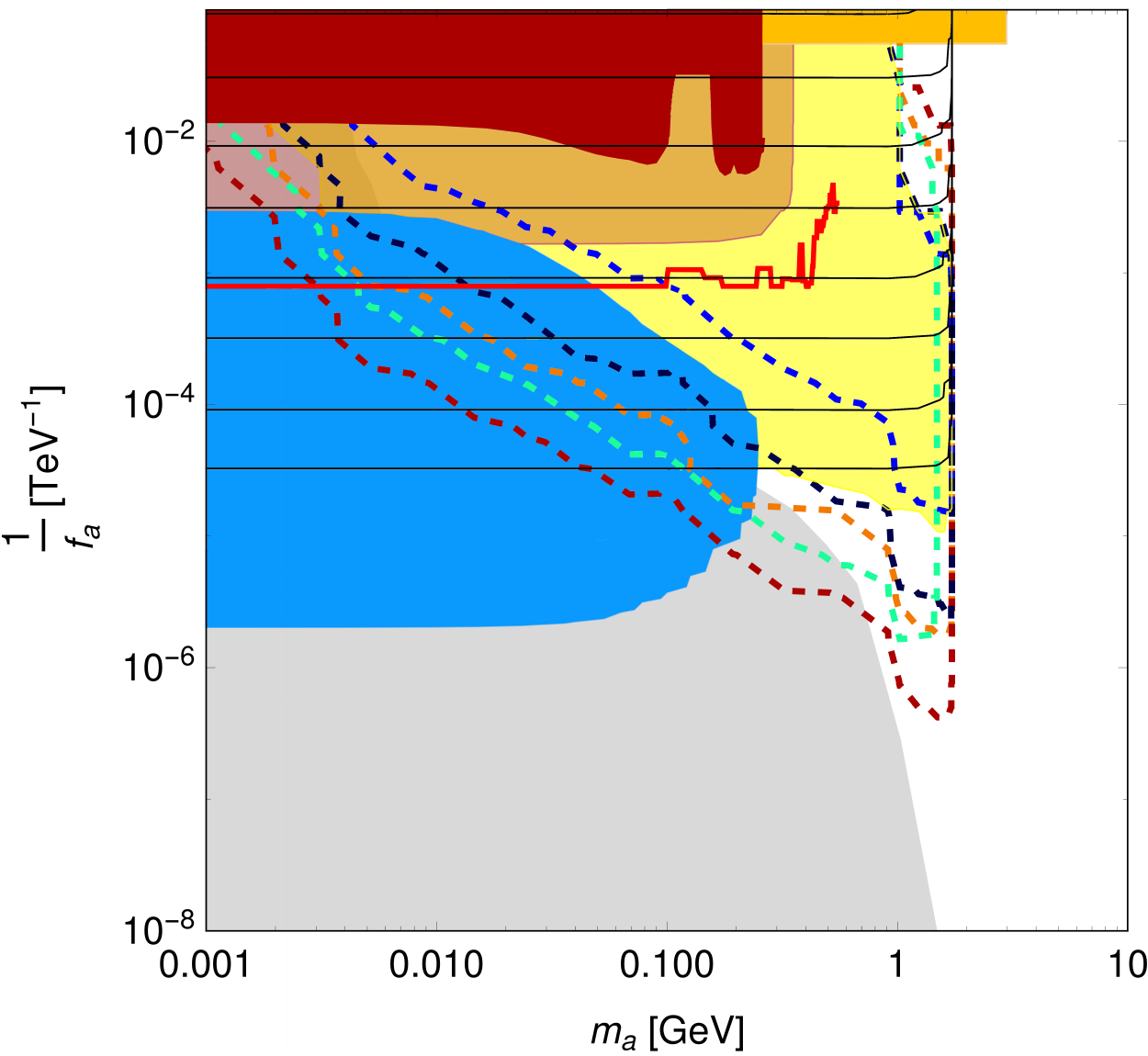}
		\caption{Same as figure~\ref{fig:ALP_results} but for the anarchic and the FN inspired models, left and right panels, respectively. Lower panels zoom in the regions where upcoming experiments are sensitive.}
		\label{fig:ALP_results2}
	\end{center}
\end{figure*}

 The resulting constraints on the parameter space of the four benchmarks models under consideration are displayed in figures~\ref{fig:ALP_results} and~\ref{fig:ALP_results2}. More specifically, we show in figure~\ref{fig:ALP_results} the different bounds for the dark-QCD motivated benchmarks, with left panels corresponding to the case $a=\pi_{D_3}$ and the right ones to $a=\pi_{D_8}$, respectively. For these two benchmarks  we further assume $\kappa_0=1$ in order to evaluate the logarithms coming from the one-loop running, which only depend on $f_a$. Still, different choices of $\kappa_0\sim\mathcal{O}(1)$ will not  significantly affect the resulting bounds shown in the figure.  On the other hand, we show in figure~\ref{fig:ALP_results2} the corresponding bounds for the anarchic (left panels) and FN (right panels) scenarios.

 In both figures, we represent in dark yellow the region of the parameter space excluded by $D$ meson mixing, whereas the bounds from exotic meson decays $B\to Ka$, $B\to \pi a$ and $K\to \pi a$ are displayed in green, blue and red, respectively. Moreover, for the sake of illustration, one can roughly interpret the expected bounds on $B\to K \nu\bar{\nu}$ from Belle II as sensitivity to $B\to Ka$, which we show in light green. Similarly, one could do the same thing in the case of $K\to \pi \nu\bar{\nu}$ from NA62, which is represented in light red. A proper recast of these two upcoming experimental results will most likely result in slightly weaker bounds. On the other hand, we show the limits arising from the recast of $D^+\to (\tau^+\to \pi^+\nu)\bar{\nu}$ as a red line.   Finally, we  show in purple the bounds resulting from $J/\psi\to a \gamma$ searches by  the CLEO collaboration.  The cosmological bounds discussed previously are shown in gray in both figures, with the constraints arising from red giant bursts and from SN1987a exhibited in lilac and light blue, respectively. 

The region where the ALP mass lies above the $D$ and $B$ meson masses is only weakly constrained by flavour observables, and is open for probes at the energy frontier, i.e. the LHC and future colliders. Below the $D$ meson mass, some viable regions remain which can be probed by the upcoming or proposed collider experiments discussed in section~\ref{sec:coll}. 
The projected bounds from SHiP and NA62 are represented by dark red and orange dashed contours, respectively. Above these lines more than three events are expected. Furthermore, the discovery lines for FASER with LHC run-3 and FASER2 at HL-LHC are displayed  in dashed light and dark blue, respectively, whereas the detection line for MATHUSLA is pictured as a turquoise dashed contour line.

In order to appreciate better the region which can be probed by these upcoming experiments, we show in the lower panels of figures~\ref{fig:ALP_results} and~\ref{fig:ALP_results2} a zoomed-in view of the area which may be reached. While FASER will mainly validate the constraints from the CHARM experiment shown in light yellow, FASER2 at the HL-LHC  as well as NA62 will probe new regions of parameter space, with SHiP and MATHUSLA covering the remaining unexplored areas below the charm mass threshold. 
%

A possible measurement of $\mathrm{Br}(D \to \pi +\rm{invisible})$ could provide a complementary test of these parts of the parameter space, and might be crucial to probe the region close to the charm mass at relatively large coupling. In the FN inspired model as well as the $\pi_D^3$ scenario this region of parameter space is not otherwise accessible, while in the $\pi_D^8$ scenario it will be probed by Belle II, and in the anarchic case it is already excluded. Similarly, the low mass region of the $\pi_D^3$ scenario is only accessible via $\mathrm{Br}(D \to \pi +\rm{invisible})$ and future NA62 measurements.
To display the discovery potential of such a measurement, we show black contour lines corresponding to values of  $\mathrm{Br}(D^\pm\to\pi^\pm \rm{invisible}) \in \{ 10^{-8},\,10^{-7},\,10^{-6},\,10^{-5},\,10^{-4},\,10^{-3},\,10^{-2},\, 10^{-1}\}$. This demonstrates that providing an experimental measurement of the exotic meson decay $D\to \pi a$ is paramount to leave no stone unturned in the quest for well motivated extensions of the SM that feature charming ALPs.

\section{Conclusions}
\label{sec:conc}

In this work we have presented several examples of models featuring charming ALPs, i.e., light pNGBs having off-diagonal couplings with SM up-quarks, and studied in detail the phenomenology associated with their low-energy EFTs. More specifically, we have studied the constraints arising from flavour experiments, astrophysics and cosmology as well as planned fixed-target and collider experiments in four benchmark models. We have shown that such scenarios have still a large unexplored parameter space. We have also demonstrated how future collider and fixed-target experiments can probe these models and that they could  be perfectly complemented by the measurement of the exotic decay $D\to \pi +\rm{invisible}$, which is currently unavailable. We thus encourage our experimental colleagues to proceed with such measurement. In the absence of dedicated searches, we have also derived bounds on the parameter space of the models by recasting three-body meson decays like $D^+\to (\tau^+\to \pi^+ \nu)\bar{\nu}$ or $B\to K/\pi\, \nu \bar{\nu}$.

The scenarios considered here  can be the low-energy EFT of several compelling UV completions. We have presented two of them: the case of a QCD-like dark sector interacting with the SM via a heavy scalar mediator with hypercharge $-2/3$, and a FN model of flavour where only RH up-quarks and a heavy scalar have non-zero charges with respect to an spontaneously broken global $U(1)$ symmetry. Some of the phenomenology studied here may change when considering the whole picture in the dark-QCD case, since there might be  a non-trivial interplay between the complete set of pNGBs in some regions of the parameter space. In the present work, we have focused on the phenomenological aspects expected to hold when singling out one of such light states. The complete dark-QCD model will be studied in an upcoming work, where a detailed study of the full charming dark sector including the possible connection with dark matter will be presented. A particularly interesting aspect of such scenarios to be studied is the collider phenomenology involving  rare top decays $t\to c a$ as well as the phenomenology of 'charming' emerging jets.

A final region of parameter space that was left unexplored in our study is that of very small ALP masses and couplings, i.e. the lower left regions of our figures. There the charming ALP would be a stable dark matter (DM) candidate. The freeze-out of such a DM candidate is excluded for $m_a\gtrsim100$~eV due to DM overproduction~ \cite{Cadamuro:2011fd} and in the whole parameter region if structure formation bounds are also taken into account~ \cite{Baumholzer:2020hvx}. 
On the other hand, if the charming ALP DM is produced via a "freeze-in" mechanism, a region of parameter space remains viable~\cite{Baumholzer:2020hvx} in principle. A more detailed exploration of this region of the charming ALP is left for future work.


\begin{acknowledgments}
We thank Felix Kahlh\"ofer and Fatih Ertas for useful feedback. AC thanks Mikael Chala, Jorge Martin-Camalich, Matthias Neubert and Robert Ziegler for fruitful discussions. AC acknowledges  funding from the European Union's Horizon 2020 research
	and innovation programme under the Marie Sk\l{}odowska-Curie grant agreement No~754446 and UGR Research and Knowledge Transfer Found -- Athenea3i. Work in Mainz was supported by the Cluster of Excellence Precision Physics, Fundamental Interactions,  and  Structure  of  Matter  (PRISMA+  EXC  2118/1)  funded  by  the  German Research Foundation (DFG) within the German Excellence Strategy (Project ID 39083149), and by grant 05H18UMCA1 of the German Federal Ministry for Education and Research (BMBF). 
\end{acknowledgments}

\appendix

\section{A dark QCD UV completion}
\label{app:dqcd}
\begin{figure}[t]
	\begin{center}
		\includegraphics[width=0.45\textwidth]{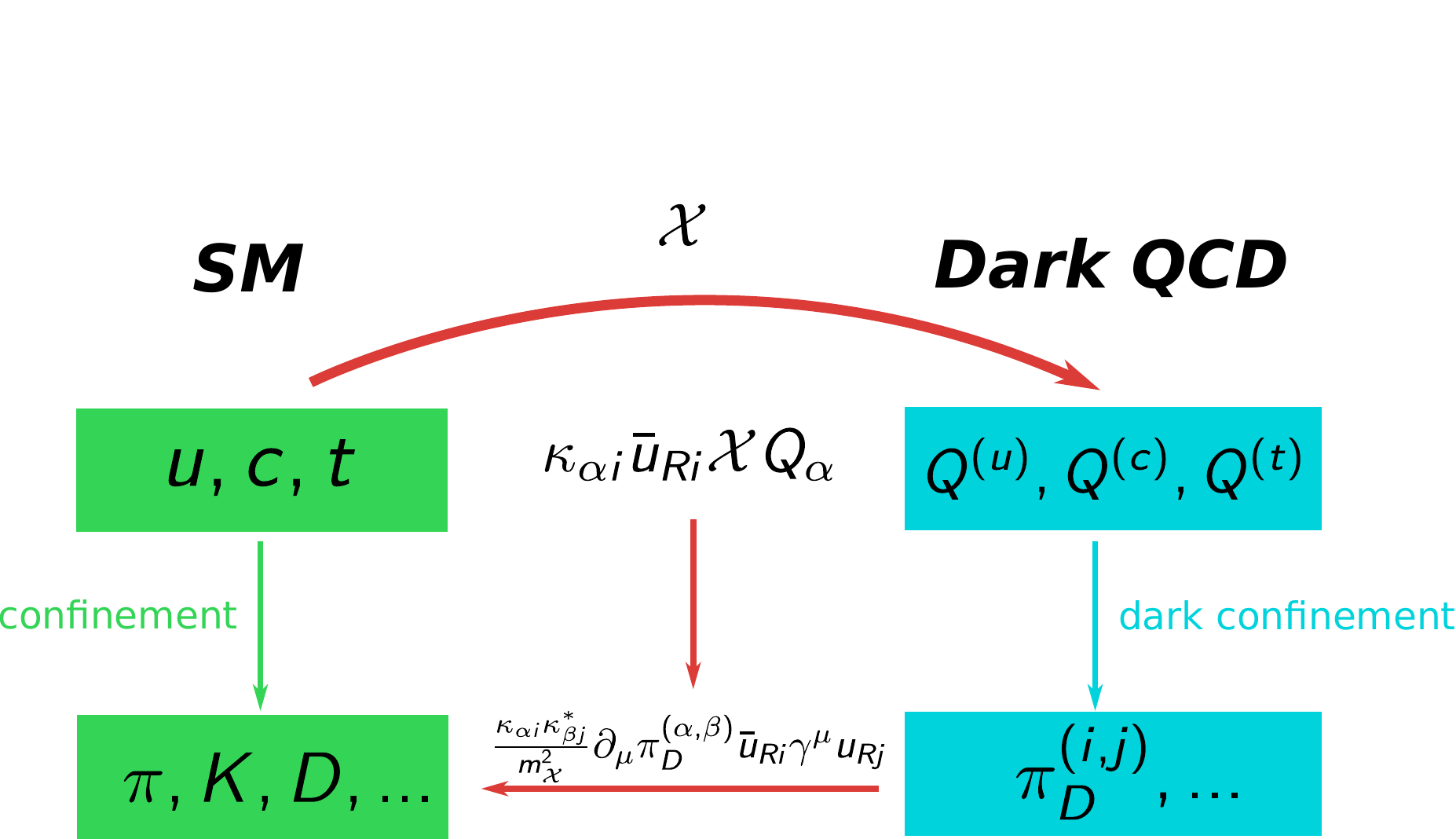}
		\caption{Schematic illustration of the high-energy and low-energy regimes of the dark QCD UV completion.}
		\label{fig:sketch}
	\end{center}
\end{figure}

One particular class of theories which can provide a UV completion to the charming ALP EFT is what can be collectively denoted by 'dark QCD', see e.g. \cite{Bai:2013xga, Schwaller:2015gea, Renner:2018fhh}. In these scenarios, one assumes that the SM is extended with a new dark QCD-like gauge group $SU(N_d)_d$ with $n_d$ Dirac fermions, $Q_{\alpha},\,\alpha=1,\ldots,n_d$, singlets of the SM gauge group and transforming in the fundamental representation of $SU(N_d)_d$. For concreteness one can assume $N_d=3=n_d$, which allows in particular the QCD-like sector to confine at a scale $\Lambda_{d\rm QCD}$. Both sectors talk to each other through the coupling to a heavy scalar mediator, $\mathcal{X}$, transforming as a $(\mathbf{3},\bar{\mathbf{3}})$ under $SU(3)\otimes SU(3)_d$, which is also charged under $SU(2)_L\otimes U(1)_Y$ as $\mathbf{1}_{-2/3}$. Such scalar mediator is naturally heavy, also in agreement with LHC bounds. A similar setup was presented first in~\cite{Bai:2013xga} but with a different assignment of hypercharge, $Y=1/3$, such that only couplings to RH down-like quarks were allowed. It was shown in~\cite{Schwaller:2015gea, Schwaller:2015gea, Renner:2018fhh} that this scenario lead to \emph{emerging jets} and its flavour phenomenology was studied in~\cite{Renner:2018fhh}. Here, we focus on a different hypercharge assignment, which leads to a distinct phenomenology. The structure of the model is sketched in figure~\ref{fig:sketch}.

The Lagrangian of the dark sector reads
\begin{align}
	\mathcal{L}_D&=-\frac{1}{4}\mathcal{G}_{d}^{\mu\nu, a}\mathcal{G}^{d,a}_{\mu\nu}+ \bar{Q}_{\alpha}i\cancel{D}Q_{\alpha}-m_{Q\alpha,\beta}\bar{Q}_{\alpha} Q_{\beta}\nonumber\\
	&+|D_{\mu}\mathcal{X}|^2-m_{\mathcal{X}}^2 |\mathcal{X}|^2-\left[\kappa_{\alpha i} \bar{u}_{Ri}\mathcal{X} Q_{\alpha} +\mathrm{h.c.}\right],
\end{align}
where we use greek (latin) indices  to denote the flavour indices in the dark (visible) QCD sector and we do not need to explicitly write the corresponding covariant derivatives. Similarly, we did not make explicit color or dark color indices with the exception of $\mathcal{G}_{\mu\nu}^{a}$, $a=1,\ldots,8$, the field-strength tensor for the dark QCD. In general, the coupling matrix $\kappa_{\alpha i}$ can be expressed as \begin{align}
	\kappa=V D U
\end{align}
where both $U$ and $V$ are $3\times 3 $ unitary matrices and $D$ is a $3\times 3$ diagonal matrix. In the case where $m_{Q\alpha\beta}=m_{Q}\delta_{\alpha\beta}$, there is a $U(3)_d$ flavour symmetry in the dark sector, which can be used to rotate away $V$. We will assume that this is the case henceforth. The matrix $U$ can be further expressed as the following product
\begin{align}
	U=U_{23}U_{13} U_{12},
\end{align}
where $U_{ij}$ are unitary rotations between the flavours $i$ and $j$. For example, $U_{12}$ reads
\begin{align}
	U_{12}=\begin{pmatrix}c_{12}&s_{12}e^{-i\delta_{12}}&0\\ -s_{12}e^{-i\delta_{12}}&c_{12}&0\\ 0&0&1 \end{pmatrix},
\end{align}
where $c_{12}=\cos\theta_{12}$, $s_{12}=\sin\theta_{12}$. Moreover, following~\cite{Agrawal:2014aoa, Renner:2018fhh} one can express $D$ as
\begin{align}
	D=\mathrm{diag}\left(\kappa_0+\kappa_1,\kappa_0+\kappa_2,\kappa_0-(\kappa_1+\kappa_2)\right),
\end{align}
where $\kappa_0\ge |\kappa_1+\kappa_2|$. For the sake of concreteness, we will consider the benchmark model defined by  
\begin{align}
	\kappa_1=\kappa_0/2,\quad \kappa_2=0,\quad  \theta_{13}=\theta_{23}=0,\quad \delta_{ij}=0,\, \forall i,j.
	\label{eq:scenarios}
\end{align}
We also assume that all this happens in a basis where the SM up Yukawa matrix is diagonal. 

If the mass of scalar mediator is much heavier than $\Lambda_{d\rm QCD}$ and $\Lambda_{\rm QCD}$, one gets in addition to the SM Lagrangian  
\begin{align}
	\mathcal{L}_{\mathrm{eff}}&=-\frac{1}{4}\mathcal{G}_{d}^{\mu\nu, a}\mathcal{G}^{d,a}_{\mu\nu}+ \bar{Q}_{\alpha}i\cancel{D}Q_{\alpha}-m_{Q}\bar{Q}_{\alpha} Q_{\alpha}\nonumber\\
	&-\frac{\kappa_{\alpha i}\kappa^{\ast}_{\beta j}}{2m_{\mathcal{X}}^2}\left(\bar{Q}_{\beta}\gamma_{\mu} P_L Q_{\alpha}\right)(\bar{u}_{Ri}\gamma^{\mu} u_{Rj}),
\end{align}
after integrating out $\mathcal{X}$ and using Fierz identities.  Similarly to QCD, in the limit $m_{Q}\to 0$ and $m_{\mathcal{X}}\to \infty$, the dark sector features a global dark chiral symmetry $SU(3)_{dL}\otimes SU(3)_{dR}$, which we assume is spontaneously  broken to its diagonal group $SU(3)_{dV}$ by the dark QCD condensate $\langle \bar{Q}_{\alpha}Q_{\beta}\rangle\propto \delta_{\alpha\beta} \Lambda_{d \rm QCD}^3$. This spontaneous symmetry breaking delivers 8 Nambu-Goldstone bosons, $\pi_{D_1},\ldots,\pi_{D_8}$, which will become pNGBs once we switch  on $m_Q$.  We will consider the case where $m_{Q}\ll \Lambda_{d\rm QCD}$ so that the  pNGBs are parametrically lighter than the rest of particles in the spectrum. With the exception of the lightest baryonic bound states carrying a conserved dark baryon number, the rest of the particles of the spectrum will undergo fast decays to dark pions. Therefore, it is a reasonable approximation to just consider  such dark pions. In the case at hand where $n_d=3=N_d$,  one can write down an effective theory for the resulting eight pNGBs along the lines of the QCD case of  pions and kaons. Using the basis of Gell-Mann matrices, $\lambda^a$, $a=1,\ldots,8$, one can write
\begin{align}
	\Pi_D&=\pi_{D_a}\frac{\lambda^{a}}{2}\\
	&=\frac{1}{2}\begin{pmatrix}\pi_{D_3}+\frac{\pi_{D_8}}{\sqrt{3}}& \pi_{D_1}-i \pi_{D_2}&\pi_{D_4}-i\pi_{D_5}\\ \pi_{D_1}+i \pi_{D_2}&-\pi_{D_3}+\frac{\pi_{D_8}}{\sqrt{3}}&\pi_{D_6}-i\pi_{D_7}\\ \pi_{D_4}+i\pi_{D_5}&\pi_{D_6}+i\pi_{D_7}& -\frac{2}{\sqrt{3}}\pi_{D_8}\end{pmatrix}.\nonumber
\end{align}
  The corresponding Goldstone matrix transforming non-linearly under $SU(3)_{dL}\otimes SU(3)_{dR}$ and linearly under $SU(3)_{dV}$ can be written as
\begin{align}
	U_D(\Pi_D)=\mathrm{exp}\left(\frac{2i}{f_d}\Pi_D\right),
\end{align}
where $f_d$ is the dark pion decay constant, in principle a free parameter. In analogy to chiral perturbation theory (ChPT) for QCD, the Lagrangian describing the dark pions in the absence of interaction with the SM is given by
\begin{align}
	\mathcal{L}_{d\rm ChPT}&=\frac{f_d^2}{4}\mathrm{Tr}\left(\partial_{\mu}U_D \partial^{\mu}U^{\dagger}_D\right)\nonumber\\
	&+\frac{f_d^2 B_{D}}{2}m_{Q}\mathrm{Tr}\left(U_D^{\dagger}+U_D\right),
\end{align}
where $B_d$ is a constant related to the dark pion mass. More precisely  
\begin{align}
	m_{\pi_{D_a}}^2=m_{\pi_D}^2=2m_{Q}B_d.
\end{align}
As one can see, they are all degenerate in mass, but small splittings will be induced by their interactions with the SM. Such radiative corrections will define new mass eigenstates   
\begin{align}
	\pi^{(1,2)}_D&=\frac{1}{\sqrt{2}}\left(\pi_{D_1}-i\pi_{D_2}\right),\\
	\pi^{(1,3)}_D&=\frac{1}{\sqrt{2}}\left(\pi_{D_3}-i\pi_{D_4}\right),\\
	\pi^{(2,3)}_D&=\frac{1}{\sqrt{2}}\left(\pi_{D_6}-i\pi_{D_7}\right),
\end{align}
with $\pi_{D_3}$ and $\pi_{D_8}$ unchanged.  

If the dark pions are light enough, $m_{\pi_D}\lesssim 4\pi f_{\pi}$, their decays are better described by ChPT for the SM quarks. The part containing only SM fields is described by 
\begin{align}
	\mathcal{L}_{\rm ChPT}=\frac{f_{\pi}^2}{4}\mathrm{Tr}\left(\partial_{\mu}U \partial^{\mu}U^{\dagger}\right)+\frac{f_{\pi}^2 B_{0}}{2}\mathrm{Tr}\left(m_{q}U^{\dagger}+Um_q^{\dagger}\right),
\end{align}
whereas the interaction with the dark QCD sector is  described by 
\begin{align}
	\mathcal{L}_{\rm ChPT}^{\rm mix}&=-\frac{f_d^2 f_{\pi}^2}{2m_{\mathcal{X}}^2}\kappa_{\alpha i}\kappa^{\ast}_{\beta j}\mathrm{Tr}\left(c_{\beta\alpha} U_D^{\dagger}	\left(\partial_{\mu}U_D\right)\right)\times\nonumber\\
	&\mathrm{Tr}\left(c_{ij}U\left(\partial^{\mu}U\right)^{\dagger}\right),\label{eq:hadmix}
\end{align}
where the projection matrices $c_{\alpha\beta}$ and $c_{ij}$ are defined as 
\begin{align}
	c_{\alpha\beta}^{mn}=\delta_{\alpha}^m\delta_{n}^{\beta}, \quad \alpha,\beta=1,2,3,\quad c_{ij}^{mn}=\delta_{i}^m\delta_{j}^n,\quad i,j=1,
\end{align}
being zero otherwise. For larger dark pion masses, one can use  quark-hadron duality~\cite{Poggio:1975af, Shifman:2000jv}, with the relevant Lagrangian  being  
\begin{align}
	\mathcal{L}_{\rm mix}=i\frac{f_d^2 }{2m_{\mathcal{X}}^2}\kappa_{\alpha i}\kappa^{\ast}_{\beta j}\mathrm{Tr}\left(c_{\beta\alpha} U_D^{\dagger}	\left(\partial_{\mu}U_D\right)\right)(\bar{u}_{Ri}\gamma^{\mu} u_{Rj}).\label{eq:parmix}
\end{align}
For the benchmark model at hand, only $\pi_{D_3}, \pi_{D_8}$ and $\pi_{D}^{(1,2)}$ decay at tree-level. The rest of dark pions, which decay through loop-induced processes, are therefore  long-lived. For this reason, we focus on the first dark pions, and in particular, on the two real ones $\pi_{D_3}$ and $\pi_{D_8}$.~\footnote{At the phenomenological level, the main difference between the couplings of these two pions  is the presence of a tree-level coupling with the RH top. The rest of pions will interpolate between these two scenarios or  be too-long lived.} In particular, the mixing terms in eqs.~\eqref{eq:hadmix} and \eqref{eq:parmix} involving  $\pi_{D_3}$ and $\pi_{D_8}$ read
\begin{align}
	\mathcal{L}_{\rm ChPT}^{\rm mix}&\supset-\frac{f_d f_{\pi}^2}{2m_{\mathcal{X}}^2}\sum_{a=3,8}\sum_{\alpha\beta}\kappa_{\alpha i}\kappa^{\ast}_{\beta j}\left(\lambda^a\right)_{\alpha\beta}\partial_{\mu}\pi_{D_a}	\times \nonumber\\
	&\mathrm{Tr}\left(c_{ij}U\left(\partial^{\mu}U\right)^{\dagger}\right)
\end{align}
and
\begin{align}
	\mathcal{L}_{\rm mix}\supset	-\frac{f_d }{2m_{\mathcal{X}}^2}\sum_{a=3,8}\sum_{\alpha\beta}\kappa_{\alpha i}\kappa^{\ast}_{\beta j}\left(\lambda^a\right)_{\alpha\beta}\partial_{\mu}\pi_{D_a}(\bar{u}_{Ri}\gamma^{\mu} u_{Rj}),
\end{align}
respectively.
Comparing these equations with the mixing terms present in eqs.~\eqref{eq:alag} and \eqref{lag:alp} for $a=\pi_{D_3}$ and $\pi_{D_8}$, respectively, we obtain $f_a=m_{\mathcal{X}}^2/f_d$ and
\begin{align}
	(c_{u_R}^{(a)})_{ij}=-\sum_{\alpha\beta}\kappa_{\alpha i}\kappa^{\ast}_{\beta j}\left(\lambda^a\right)_{\alpha\beta},\qquad a=3,8.
\end{align}
More explicitly,
\begin{align}
	c_{u_R}^{(3)}&=\frac{\kappa_0^2}{4}\begin{pmatrix}4s_{12}^2-9c_{12}^2& -13 c_{12}s_{12}&0\\-13c_{12}s_{12} & 4c_{12}^2-9s_{12}^2  &0\\ 0&0&0\end{pmatrix},\\
		c_{u_R}^{(8)}&=\frac{-\kappa_0^2}{4\sqrt{3}}\begin{pmatrix}4s_{12}^2+9c_{12}^2& 5c_{12}s_{12}&0\\5c_{12}s_{12} & 4c_{12}^2+9s_{12}^2 &0\\ 0&0&-2\end{pmatrix}.
\end{align}
\section{A Froggatt-Nielsen UV completion}
\label{app:fn}
Another motivation for the ALPs considered here corresponds to what is  generically known by the name of flavons or familions~(see \cite{Davidson:1981zd, Wilczek:1982rv, Reiss:1982sq, Berezhiani:1990wn, Berezhiani:1990jj, Feng:1997tn} and e.g.~\cite{Albrecht:2010xh, Bauer:2016rxs, Calibbi:2016hwq, Ema:2016ops, Ema:2018abj, Heikinheimo:2018luc, Bonnefoy:2019lsn, Egana-Ugrinovic:2019wzj, Bonnefoy:2020llz, Alonso-Alvarez:2021ett} for more recent implementations), pNGBs of some spontaneously broken flavour symmetry, which may be anomalous, and that generically feature flavour-violating couplings to quarks or leptons. One particular setup leading to the scenario we have in mind, i.e., the effective Lagrangian~(\ref{eq:alag}) in addition to $c_{H}=0$, is given by FN models when only RH up-quarks have non-zero charges. Specifically, one considers a global $U(1)$ flavour symmetry, spontaneously broken by the vacuum expectation value of some extra scalar $\langle S\rangle =f_a$, where
\begin{align}
	S=\frac{1}{\sqrt{2}}(f_a+s)e^{i a/f_a},
\end{align}
and has charge $-1$ under this new $U(1)$. If only $u_{Ri}$ are charged under such global symmetry, with charges $n_{i}^{u}$, Yukawa couplings for up-quarks will be higher-dimensional
\begin{align}
	\mathcal{L}\supset -(y_u)_{ij}\left( \frac{S}{\Lambda}\right)^{n_j^u} q_{Li} \tilde{H} u_{Rj}+\mathrm{h.c.},
\end{align}
where one typically assumes that $f_a<\Lambda$. At the end of the day, such term in the Lagrangian will generate interactions like~(\ref{eq:yuks})
\begin{align}
	-\frac{ia}{f_a}\bar{q}_{Li} \tilde{H} u_{Rj} n_{j}^u=	-\frac{ia}{f_a}\bar{q}_{Li} \tilde{H} u_{Rj} (Y_u)_{ij}n_{j}^u,
	\label{eq:yukef}
\end{align}
where $Y_u$ is the effective up Yukawa matrix, 
\begin{align}
	(Y_u)_{ij}=(y_{u})_{ij}\left(\frac{f_a}{\Lambda}\right)^{n_{j}^u}.
\end{align}
If we assume that $(y_{u})_{ij}=\mathcal{O}(1)$  and take $f_a/\Lambda=\epsilon\sim m_c/m_t$, we can get the correct up quark masses by choosing $n_u=(2,1,0)$
since
\begin{align}
	(m_u,m_c,m_t)\sim \frac{1}{\sqrt{2}} v \, (\epsilon^2,\epsilon,1)
\end{align}
and
\begin{align}
	\frac{m_u}{m_t}\sim \epsilon^2,\qquad \frac{m_c}{m_t}\sim \epsilon.
\end{align}
In this case, we can diagonalize $Y_u$ by making 
\begin{align}
	u_R\to U_R^u u_R,\qquad u_L\to U_L^u u_L,
	\end{align}
	with
	\begin{align}
		U_R^u\sim \begin{pmatrix}1&\epsilon&\epsilon^2\\ \epsilon&1&\epsilon\\ \epsilon^2& \epsilon&1 \end{pmatrix},\qquad (U_L^u)_{ij}\sim\mathcal{O}(1). 
	\end{align}
This leads, after going to the basis where $Y_u$ is diagonal to
\begin{align}
	c_{u_R}\sim\begin{pmatrix}2&3\epsilon&3\epsilon^2\\ 3\epsilon&1&\epsilon\\3 \epsilon^2& \epsilon & \epsilon^2\end{pmatrix}.
\end{align}
\phantom{This sentence is here to fix the layout.}
\section{ALP couplings to nucleons}
\label{app:cann}
The leading order ALP couplings with nucleons  can be read from the following Lagrangian~\cite{Kaplan:1985dv, Srednicki:1985xd, Georgi:1986df, Chang:1993gm}
\begin{widetext}
\begin{align}
	\mathcal{L}_{\rm int}&=	\left(\frac{\partial_{\mu} a}{4f_a}\right)\Big\{\mathrm{Tr}\left((\hat{c}+\varkappa_q c_{g})\lambda^a\right)\left(F\, \mathrm{Tr}\left(\bar{B}\gamma^{\mu}\gamma_5\left[\lambda^a,B\right]\right)+D\, \mathrm{Tr}\left(\bar{B}\,\gamma^{\mu}\gamma_5\left\{\lambda^a,B\right\}\right) \right) \nonumber\\
	+& \frac{1}{3}\mathrm{Tr}\left( \hat{c}+\varkappa_q c_{g}\right)\, S\, \mathrm{Tr}\left(\bar{B}\,\gamma^{\mu}\gamma_5 B\right)+\mathrm{Tr}\left( (\hat{c}+\varkappa_q c_{g})\lambda^a\right) \mathrm{Tr}\left(\bar{B}\,\gamma^\mu \left[\lambda^a,B\right]\right)\Big\}
\end{align}
\end{widetext}
where $B$ is the baryon matrix
\begin{align}
	B=\frac{1}{\sqrt{2}}B^a\lambda^a=\begin{pmatrix}\frac{\Sigma^0}{\sqrt{2}}+\frac{\Lambda}{\sqrt{6}}& \Sigma^+& p\\ \Sigma^-& -\frac{\Sigma^0}{\sqrt{2}}+\frac{\Lambda}{\sqrt{6}}& n\\ \Xi^{-} & \Xi^0& -\frac{2\Lambda}{\sqrt{6}} \end{pmatrix},
\end{align}
and the axial-vector coupling constants $F$ and $D$ are defined by
\begin{align}
	\langle B^{\prime}_i| J_j^{(8)}| B_k\rangle = i f_{ijk}F+d_{ijk}D,
\end{align}
with $J_j^{(8)}$ the weak axial-vector hadronic current, transforming as an $SU(3)$ octet,  $f_{ijk}$ the totally antisymmetric structure constants of $SU(3)$ and $d_{ijk}$ the totally symmetric ones. 
On the other hand, $S$ is defined by the singlet current which can be renormalized independently. $D$ and $F$ can be determined by hyperon semileptonic decays~\cite{Cabibbo:2003cu}, leading to $F=0.463\pm0.008$, $D=0.804\pm 0.008$.  On the other hand $S\approx 0.13\pm 0.2$~\cite{Jaffe:1989jz}.
We are particularly interested in the $a \bar{N} N$ couplings, with $N=p,n$, which read
\begin{widetext}
\begin{align}
\mathcal{L}_{\rm int}\supset\frac{1}{12}\frac{\partial_{\mu} a}{f_a}\left(c_{u_R}\right)_{11}\Big( \big[S-4D\big](\bar{n}\gamma^{\mu}\gamma_5 n)+\big[2D+6F+S\big](\bar{p}\gamma^{\mu}\gamma_5p)+6\bar{p}\gamma^{\mu} p \Big)\end{align}
\end{widetext}
The last term in the equation above is a total derivative which can be neglected. In this case
\begin{align}
	\mathcal{L}_{aNN}=\sum_{N=p,n}\frac{\partial_{\mu}a}{2f_a}c_{aNN}\bar{N}\gamma^{\mu}\gamma_5 N
\end{align}
with
\begin{align}
	c_{app}&=(c_{u_R})_{11}\left(F+\frac{1}{3}D+\frac{1}{6}S\right)=(c_{u_R})_{11}\left(0.75\pm0.03\right),
\end{align}
\begin{align}
	c_{ann}&=(c_{u_R})_{11}\left(\frac{1}{6}S-\frac{2}{3}D\right)=(c_{u_R})_{11}\left(-0.51\pm0.03\right).
\end{align}
\vspace{0.1cm}

\bibliography{refs}{}

\providecommand{\href}[2]{#2}\begingroup\raggedright\begin{thebibliography}{100}

\bibitem{Craig:2015pha}
N.~Craig, A.~Katz, M.~Strassler and R.~Sundrum, \emph{{Naturalness in the Dark
  at the LHC}}, \href{https://doi.org/10.1007/JHEP07(2015)105}{\emph{JHEP}
  {\bfseries 07} (2015) 105}
  [\href{https://arxiv.org/abs/1501.05310}{{\ttfamily 1501.05310}}].

\bibitem{Agrawal:2014aoa}
P.~Agrawal, M.~Blanke and K.~Gemmler, \emph{{Flavored dark matter beyond
  Minimal Flavor Violation}},
  \href{https://doi.org/10.1007/JHEP10(2014)072}{\emph{JHEP} {\bfseries 10}
  (2014) 072} [\href{https://arxiv.org/abs/1405.6709}{{\ttfamily 1405.6709}}].

\bibitem{Batell:2011tc}
B.~Batell, J.~Pradler and M.~Spannowsky, \emph{{Dark Matter from Minimal Flavor
  Violation}}, \href{https://doi.org/10.1007/JHEP08(2011)038}{\emph{JHEP}
  {\bfseries 08} (2011) 038} [\href{https://arxiv.org/abs/1105.1781}{{\ttfamily
  1105.1781}}].

\bibitem{Calibbi:2015sfa}
L.~Calibbi, A.~Crivellin and B.~Zald\'\i{}var, \emph{{Flavor portal to dark
  matter}}, \href{https://doi.org/10.1103/PhysRevD.92.016004}{\emph{Phys. Rev.
  D} {\bfseries 92} (2015) 016004}
  [\href{https://arxiv.org/abs/1501.07268}{{\ttfamily 1501.07268}}].

\bibitem{Renner:2018fhh}
S.~Renner and P.~Schwaller, \emph{{A flavoured dark sector}},
  \href{https://doi.org/10.1007/JHEP08(2018)052}{\emph{JHEP} {\bfseries 08}
  (2018) 052} [\href{https://arxiv.org/abs/1803.08080}{{\ttfamily
  1803.08080}}].

\bibitem{Mies:2020mzw}
H.~Mies, C.~Scherb and P.~Schwaller, \emph{{Collider constraints on dark
  mediators}},  \href{https://arxiv.org/abs/2011.13990}{{\ttfamily
  2011.13990}}.

\bibitem{Blanke:2017tnb}
M.~Blanke and S.~Kast, \emph{{Top-Flavoured Dark Matter in Dark Minimal Flavour
  Violation}}, \href{https://doi.org/10.1007/JHEP05(2017)162}{\emph{JHEP}
  {\bfseries 05} (2017) 162}
  [\href{https://arxiv.org/abs/1702.08457}{{\ttfamily 1702.08457}}].

\bibitem{Jubb:2017rhm}
T.~Jubb, M.~Kirk and A.~Lenz, \emph{{Charming Dark Matter}},
  \href{https://doi.org/10.1007/JHEP12(2017)010}{\emph{JHEP} {\bfseries 12}
  (2017) 010} [\href{https://arxiv.org/abs/1709.01930}{{\ttfamily
  1709.01930}}].

\bibitem{Blanke:2020bsf}
M.~Blanke, P.~Pani, G.~Polesello and G.~Rovelli, \emph{{Single-top final states
  as a probe of top-flavoured dark matter models at the LHC}},
  \href{https://arxiv.org/abs/2010.10530}{{\ttfamily 2010.10530}}.

\bibitem{Strassler:2006im}
M.~J. Strassler and K.~M. Zurek, \emph{{Echoes of a hidden valley at hadron
  colliders}},
  \href{https://doi.org/10.1016/j.physletb.2007.06.055}{\emph{Phys. Lett. B}
  {\bfseries 651} (2007) 374}
  [\href{https://arxiv.org/abs/hep-ph/0604261}{{\ttfamily hep-ph/0604261}}].

\bibitem{Bai:2013xga}
Y.~Bai and P.~Schwaller, \emph{{Scale of dark QCD}},
  \href{https://doi.org/10.1103/PhysRevD.89.063522}{\emph{Phys. Rev.}
  {\bfseries D89} (2014) 063522}
  [\href{https://arxiv.org/abs/1306.4676}{{\ttfamily 1306.4676}}].

\bibitem{Schwaller:2015gea}
P.~Schwaller, D.~Stolarski and A.~Weiler, \emph{{Emerging Jets}},
  \href{https://doi.org/10.1007/JHEP05(2015)059}{\emph{JHEP} {\bfseries 05}
  (2015) 059} [\href{https://arxiv.org/abs/1502.05409}{{\ttfamily
  1502.05409}}].

\bibitem{Cheng:2019yai}
H.-C. Cheng, L.~Li, E.~Salvioni and C.~B. Verhaaren, \emph{{Light Hidden Mesons
  through the Z Portal}},
  \href{https://doi.org/10.1007/JHEP11(2019)031}{\emph{JHEP} {\bfseries 11}
  (2019) 031} [\href{https://arxiv.org/abs/1906.02198}{{\ttfamily
  1906.02198}}].

\bibitem{Froggatt:1978nt}
C.~D. Froggatt and H.~B. Nielsen, \emph{{Hierarchy of Quark Masses, Cabibbo
  Angles and CP Violation}},
  \href{https://doi.org/10.1016/0550-3213(79)90316-X}{\emph{Nucl. Phys.}
  {\bfseries B147} (1979) 277}.

\bibitem{Jaeckel:2015jla}
J.~Jaeckel and M.~Spannowsky, \emph{{Probing MeV to 90 GeV axion-like particles
  with LEP and LHC}},
  \href{https://doi.org/10.1016/j.physletb.2015.12.037}{\emph{Phys. Lett. B}
  {\bfseries 753} (2016) 482}
  [\href{https://arxiv.org/abs/1509.00476}{{\ttfamily 1509.00476}}].

\bibitem{Brivio:2017ije}
I.~Brivio, M.~B. Gavela, L.~Merlo, K.~Mimasu, J.~M. No, R.~del Rey et~al.,
  \emph{{ALPs Effective Field Theory and Collider Signatures}},
  \href{https://doi.org/10.1140/epjc/s10052-017-5111-3}{\emph{Eur. Phys. J.}
  {\bfseries C77} (2017) 572}
  [\href{https://arxiv.org/abs/1701.05379}{{\ttfamily 1701.05379}}].

\bibitem{Bellazzini:2017neg}
B.~Bellazzini, A.~Mariotti, D.~Redigolo, F.~Sala and J.~Serra, \emph{{$R$-axion
  at colliders}},
  \href{https://doi.org/10.1103/PhysRevLett.119.141804}{\emph{Phys. Rev. Lett.}
  {\bfseries 119} (2017) 141804}
  [\href{https://arxiv.org/abs/1702.02152}{{\ttfamily 1702.02152}}].

\bibitem{Bauer:2017ris}
M.~Bauer, M.~Neubert and A.~Thamm, \emph{{Collider Probes of Axion-Like
  Particles}}, \href{https://doi.org/10.1007/JHEP12(2017)044}{\emph{JHEP}
  {\bfseries 12} (2017) 044}
  [\href{https://arxiv.org/abs/1708.00443}{{\ttfamily 1708.00443}}].

\bibitem{Knapen:2017ebd}
S.~Knapen, T.~Lin, H.~K. Lou and T.~Melia, \emph{{LHC limits on axion-like
  particles from heavy-ion collisions}},
  \href{https://doi.org/10.23727/CERN-Proceedings-2018-001.65}{\emph{CERN
  Proc.} {\bfseries 1} (2018) 65}
  [\href{https://arxiv.org/abs/1709.07110}{{\ttfamily 1709.07110}}].

\bibitem{Bauer:2018uxu}
M.~Bauer, M.~Heiles, M.~Neubert and A.~Thamm, \emph{{Axion-Like Particles at
  Future Colliders}},
  \href{https://doi.org/10.1140/epjc/s10052-019-6587-9}{\emph{Eur. Phys. J. C}
  {\bfseries 79} (2019) 74} [\href{https://arxiv.org/abs/1808.10323}{{\ttfamily
  1808.10323}}].

\bibitem{Aloni:2018vki}
D.~Aloni, Y.~Soreq and M.~Williams, \emph{{Coupling QCD-Scale Axionlike
  Particles to Gluons}},
  \href{https://doi.org/10.1103/PhysRevLett.123.031803}{\emph{Phys. Rev. Lett.}
  {\bfseries 123} (2019) 031803}
  [\href{https://arxiv.org/abs/1811.03474}{{\ttfamily 1811.03474}}].

\bibitem{Batell:2009jf}
B.~Batell, M.~Pospelov and A.~Ritz, \emph{{Multi-lepton Signatures of a Hidden
  Sector in Rare B Decays}},
  \href{https://doi.org/10.1103/PhysRevD.83.054005}{\emph{Phys. Rev. D}
  {\bfseries 83} (2011) 054005}
  [\href{https://arxiv.org/abs/0911.4938}{{\ttfamily 0911.4938}}].

\bibitem{Kamenik:2011vy}
J.~F. Kamenik and C.~Smith, \emph{{FCNC portals to the dark sector}},
  \href{https://doi.org/10.1007/JHEP03(2012)090}{\emph{JHEP} {\bfseries 03}
  (2012) 090} [\href{https://arxiv.org/abs/1111.6402}{{\ttfamily 1111.6402}}].

\bibitem{Gavela:2019wzg}
M.~B. Gavela, R.~Houtz, P.~Quilez, R.~Del~Rey and O.~Sumensari, \emph{{Flavor
  constraints on electroweak ALP couplings}},
  \href{https://doi.org/10.1140/epjc/s10052-019-6889-y}{\emph{Eur. Phys. J.}
  {\bfseries C79} (2019) 369}
  [\href{https://arxiv.org/abs/1901.02031}{{\ttfamily 1901.02031}}].

\bibitem{Bauer:2019gfk}
M.~Bauer, M.~Neubert, S.~Renner, M.~Schnubel and A.~Thamm, \emph{{Axionlike
  Particles, Lepton-Flavor Violation, and a New Explanation of $a_\mu$ and
  $a_e$}}, \href{https://doi.org/10.1103/PhysRevLett.124.211803}{\emph{Phys.
  Rev. Lett.} {\bfseries 124} (2020) 211803}
  [\href{https://arxiv.org/abs/1908.00008}{{\ttfamily 1908.00008}}].

\bibitem{Cornella:2019uxs}
C.~Cornella, P.~Paradisi and O.~Sumensari, \emph{{Hunting for ALPs with Lepton
  Flavor Violation}},
  \href{https://doi.org/10.1007/JHEP01(2020)158}{\emph{JHEP} {\bfseries 01}
  (2020) 158} [\href{https://arxiv.org/abs/1911.06279}{{\ttfamily
  1911.06279}}].

\bibitem{Calibbi:2020jvd}
L.~Calibbi, D.~Redigolo, R.~Ziegler and J.~Zupan, \emph{{Looking forward to
  Lepton-flavor-violating ALPs}},
  \href{https://arxiv.org/abs/2006.04795}{{\ttfamily 2006.04795}}.

\bibitem{Choi:2017gpf}
K.~Choi, S.~H. Im, C.~B. Park and S.~Yun, \emph{{Minimal Flavor Violation with
  Axion-like Particles}},
  \href{https://doi.org/10.1007/JHEP11(2017)070}{\emph{JHEP} {\bfseries 11}
  (2017) 070} [\href{https://arxiv.org/abs/1708.00021}{{\ttfamily
  1708.00021}}].

\bibitem{MartinCamalich:2020dfe}
J.~Martin~Camalich, M.~Pospelov, P.~N.~H. Vuong, R.~Ziegler and J.~Zupan,
  \emph{{Quark Flavor Phenomenology of the QCD Axion}},
  \href{https://doi.org/10.1103/PhysRevD.102.015023}{\emph{Phys. Rev. D}
  {\bfseries 102} (2020) 015023}
  [\href{https://arxiv.org/abs/2002.04623}{{\ttfamily 2002.04623}}].

\bibitem{Chala:2020wvs}
M.~Chala, G.~Guedes, M.~Ramos and J.~Santiago, \emph{{Running in the ALPs}},
  \href{https://arxiv.org/abs/2012.09017}{{\ttfamily 2012.09017}}.

\bibitem{Bauer:2020jbp}
M.~Bauer, M.~Neubert, S.~Renner, M.~Schnubel and A.~Thamm, \emph{{The
  Low-Energy Effective Theory of Axions and ALPs}},
  \href{https://arxiv.org/abs/2012.12272}{{\ttfamily 2012.12272}}.

\bibitem{Marciano:2016yhf}
W.~J. Marciano, A.~Masiero, P.~Paradisi and M.~Passera, \emph{{Contributions of
  axionlike particles to lepton dipole moments}},
  \href{https://doi.org/10.1103/PhysRevD.94.115033}{\emph{Phys. Rev. D}
  {\bfseries 94} (2016) 115033}
  [\href{https://arxiv.org/abs/1607.01022}{{\ttfamily 1607.01022}}].

\bibitem{DiLuzio:2020oah}
L.~Di~Luzio, R.~Gr\"ober and P.~Paradisi, \emph{{Hunting for the CP violating
  ALP}},  \href{https://arxiv.org/abs/2010.13760}{{\ttfamily 2010.13760}}.

\bibitem{Georgi:1986df}
H.~Georgi, D.~B. Kaplan and L.~Randall, \emph{{Manifesting the Invisible Axion
  at Low-energies}},
  \href{https://doi.org/10.1016/0370-2693(86)90688-X}{\emph{Phys. Lett.}
  {\bfseries 169B} (1986) 73}.

\bibitem{Choi:1986zw}
K.~Choi, K.~Kang and J.~E. Kim, \emph{{Effects of $\eta^\prime$ in Low-energy
  Axion Physics}},
  \href{https://doi.org/10.1016/0370-2693(86)91273-6}{\emph{Phys. Lett.}
  {\bfseries B181} (1986) 145}.

\bibitem{Bardeen:1986yb}
W.~A. Bardeen, R.~D. Peccei and T.~Yanagida, \emph{{Constraints on Variant
  Axion Models}},
  \href{https://doi.org/10.1016/0550-3213(87)90003-4}{\emph{Nucl. Phys.}
  {\bfseries B279} (1987) 401}.

\bibitem{Krauss:1986bq}
L.~M. Krauss and M.~B. Wise, \emph{{Constraints on Shortlived Axions From the
  Decay $\pi^+ \to e^+ e^- e^+$ Neutrino}},
  \href{https://doi.org/10.1016/0370-2693(86)90201-7}{\emph{Phys. Lett.}
  {\bfseries B176} (1986) 483}.

\bibitem{Bauer:2015fxa}
M.~Bauer, M.~Carena and K.~Gemmler, \emph{{Flavor from the Electroweak Scale}},
  \href{https://doi.org/10.1007/JHEP11(2015)016}{\emph{JHEP} {\bfseries 11}
  (2015) 016} [\href{https://arxiv.org/abs/1506.01719}{{\ttfamily
  1506.01719}}].

\bibitem{Bauer:2015kzy}
M.~Bauer, M.~Carena and K.~Gemmler, \emph{{Creating the fermion mass
  hierarchies with multiple Higgs bosons}},
  \href{https://doi.org/10.1103/PhysRevD.94.115030}{\emph{Phys. Rev.}
  {\bfseries D94} (2016) 115030}
  [\href{https://arxiv.org/abs/1512.03458}{{\ttfamily 1512.03458}}].

\bibitem{Bauer:2017cov}
M.~Bauer, M.~Carena and A.~Carmona, \emph{{Higgs Pair Production as a Signal of
  Enhanced Yukawa Couplings}},
  \href{https://doi.org/10.1103/PhysRevLett.121.021801}{\emph{Phys. Rev. Lett.}
  {\bfseries 121} (2018) 021801}
  [\href{https://arxiv.org/abs/1801.00363}{{\ttfamily 1801.00363}}].

\bibitem{Ciuchini:1998ix}
M.~Ciuchini et~al., \emph{{Delta M(K) and epsilon(K) in SUSY at the
  next-to-leading order}},
  \href{https://doi.org/10.1088/1126-6708/1998/10/008}{\emph{JHEP} {\bfseries
  10} (1998) 008} [\href{https://arxiv.org/abs/hep-ph/9808328}{{\ttfamily
  hep-ph/9808328}}].

\bibitem{Zyla:2020zbs}
{\scshape Particle Data Group} collaboration, \emph{{Review of Particle
  Physics}}, \href{https://doi.org/10.1093/ptep/ptaa104}{\emph{PTEP} {\bfseries
  2020} (2020) 083C01}.

\bibitem{Bazavov:2017weg}
A.~Bazavov et~al., \emph{{Short-distance matrix elements for $D^0$-meson mixing
  for $N_f=2+1$ lattice QCD}},
  \href{https://doi.org/10.1103/PhysRevD.97.034513}{\emph{Phys. Rev. D}
  {\bfseries 97} (2018) 034513}
  [\href{https://arxiv.org/abs/1706.04622}{{\ttfamily 1706.04622}}].

\bibitem{Golowich:2009ii}
E.~Golowich, J.~Hewett, S.~Pakvasa and A.~A. Petrov, \emph{{Relating D0-anti-D0
  Mixing and D0 ---\ensuremath{>} l+ l- with New Physics}},
  \href{https://doi.org/10.1103/PhysRevD.79.114030}{\emph{Phys. Rev. D}
  {\bfseries 79} (2009) 114030}
  [\href{https://arxiv.org/abs/0903.2830}{{\ttfamily 0903.2830}}].

\bibitem{Amhis:2019ckw}
{\scshape HFLAV} collaboration, \emph{{Averages of $b$-hadron, $c$-hadron, and
  $\tau$-lepton properties as of 2018}},
  \href{https://arxiv.org/abs/1909.12524}{{\ttfamily 1909.12524}}.

\bibitem{Lubicz:2017syv}
{\scshape ETM} collaboration, \emph{{Scalar and vector form factors of $D \to
  \pi(K) \ell \nu$ decays with $N_f=2+1+1$ twisted fermions}},
  \href{https://doi.org/10.1103/PhysRevD.96.054514}{\emph{Phys. Rev. D}
  {\bfseries 96} (2017) 054514}
  [\href{https://arxiv.org/abs/1706.03017}{{\ttfamily 1706.03017}}].

\bibitem{Bailey:2015dka}
J.~A. Bailey et~al., \emph{{$B\to Kl^+l^-$ Decay Form Factors from Three-Flavor
  Lattice QCD}}, \href{https://doi.org/10.1103/PhysRevD.93.025026}{\emph{Phys.
  Rev. D} {\bfseries 93} (2016) 025026}
  [\href{https://arxiv.org/abs/1509.06235}{{\ttfamily 1509.06235}}].

\bibitem{Gubernari:2018wyi}
N.~Gubernari, A.~Kokulu and D.~van Dyk, \emph{{$B\to P$ and $B\to V$ Form
  Factors from $B$-Meson Light-Cone Sum Rules beyond Leading Twist}},
  \href{https://doi.org/10.1007/JHEP01(2019)150}{\emph{JHEP} {\bfseries 01}
  (2019) 150} [\href{https://arxiv.org/abs/1811.00983}{{\ttfamily
  1811.00983}}].

\bibitem{Carrasco:2016kpy}
N.~Carrasco, P.~Lami, V.~Lubicz, L.~Riggio, S.~Simula and C.~Tarantino,
  \emph{{$K \to \pi$ semileptonic form factors with $N_f=2+1+1$ twisted mass
  fermions}}, \href{https://doi.org/10.1103/PhysRevD.93.114512}{\emph{Phys.
  Rev. D} {\bfseries 93} (2016) 114512}
  [\href{https://arxiv.org/abs/1602.04113}{{\ttfamily 1602.04113}}].

\bibitem{Eisenstein:2008aa}
{\scshape CLEO} collaboration, \emph{{Precision Measurement of B(D+
  ---\ensuremath{>} mu+ nu) and the Pseudoscalar Decay Constant f(D+)}},
  \href{https://doi.org/10.1103/PhysRevD.78.052003}{\emph{Phys. Rev. D}
  {\bfseries 78} (2008) 052003}
  [\href{https://arxiv.org/abs/0806.2112}{{\ttfamily 0806.2112}}].

\bibitem{Ablikim:2019rpl}
{\scshape BESIII} collaboration, \emph{{Observation of the leptonic decay $D^+
  \to \tau^+ \nu_\tau$}},
  \href{https://doi.org/10.1103/PhysRevLett.123.211802}{\emph{Phys. Rev. Lett.}
  {\bfseries 123} (2019) 211802}
  [\href{https://arxiv.org/abs/1908.08877}{{\ttfamily 1908.08877}}].

\bibitem{Kamenik:2009kc}
J.~F. Kamenik and C.~Smith, \emph{{Tree-level contributions to the rare decays
  $B^+ \to \pi^+ \nu \bar{\nu}$, $B^+\to K^+ \nu \bar{\nu}$, and $B^+ \to
  K^{\ast +} \nu\bar{\nu}$ in the Standard Model}},
  \href{https://doi.org/10.1016/j.physletb.2009.09.041}{\emph{Phys. Lett. B}
  {\bfseries 680} (2009) 471}
  [\href{https://arxiv.org/abs/0908.1174}{{\ttfamily 0908.1174}}].

\bibitem{Brun:1997pa}
R.~Brun and F.~Rademakers, \emph{{ROOT: An object oriented data analysis
  framework}}, \href{https://doi.org/10.1016/S0168-9002(97)00048-X}{\emph{Nucl.
  Instrum. Meth. A} {\bfseries 389} (1997) 81}.

\bibitem{Read:2002hq}
A.~L. Read, \emph{{Presentation of search results: The CL(s) technique}},
  \href{https://doi.org/10.1088/0954-3899/28/10/313}{\emph{J. Phys. G}
  {\bfseries 28} (2002) 2693}.

\bibitem{Adler:2008zza}
{\scshape E949, E787} collaboration, \emph{{Measurement of the K+
  --\ensuremath{>} pi+ nu nu branching ratio}},
  \href{https://doi.org/10.1103/PhysRevD.77.052003}{\emph{Phys. Rev. D}
  {\bfseries 77} (2008) 052003}
  [\href{https://arxiv.org/abs/0709.1000}{{\ttfamily 0709.1000}}].

\bibitem{Ammar:2001gi}
{\scshape CLEO} collaboration, \emph{{Search for the familon via B+-
  ---\ensuremath{>} pi+- X0, B+- ---\ensuremath{>} K+- X0, and B0
  ---\ensuremath{>} K0(S)X0 decays}},
  \href{https://doi.org/10.1103/PhysRevLett.87.271801}{\emph{Phys. Rev. Lett.}
  {\bfseries 87} (2001) 271801}
  [\href{https://arxiv.org/abs/hep-ex/0106038}{{\ttfamily hep-ex/0106038}}].

\bibitem{CortinaGil:2020fcx}
{\scshape NA62} collaboration, \emph{{Search for a feebly interacting particle
  $X$ in the decay $K^{+}\rightarrow\pi^{+}X$}},
  \href{https://arxiv.org/abs/2011.11329}{{\ttfamily 2011.11329}}.

\bibitem{Lees:2013kla}
{\scshape BaBar} collaboration, \emph{{Search for $B \to K^{(*)} \nu \overline
  \nu$ and invisible quarkonium decays}},
  \href{https://doi.org/10.1103/PhysRevD.87.112005}{\emph{Phys. Rev. D}
  {\bfseries 87} (2013) 112005}
  [\href{https://arxiv.org/abs/1303.7465}{{\ttfamily 1303.7465}}].

\bibitem{Aubert:2004ws}
{\scshape BaBar} collaboration, \emph{{A search for the decay $B^+ \to K^+ \nu
  \bar{\nu}$}},
  \href{https://doi.org/10.1103/PhysRevLett.94.101801}{\emph{Phys. Rev. Lett.}
  {\bfseries 94} (2005) 101801}
  [\href{https://arxiv.org/abs/hep-ex/0411061}{{\ttfamily hep-ex/0411061}}].

\bibitem{Buras:2014fpa}
A.~J. Buras, J.~Girrbach-Noe, C.~Niehoff and D.~M. Straub, \emph{{$ B\to
  {K}^{\left(\ast \right)}\nu \overline{\nu} $ decays in the Standard Model and
  beyond}}, \href{https://doi.org/10.1007/JHEP02(2015)184}{\emph{JHEP}
  {\bfseries 02} (2015) 184} [\href{https://arxiv.org/abs/1409.4557}{{\ttfamily
  1409.4557}}].

\bibitem{Kou:2018nap}
{\scshape Belle-II} collaboration, \emph{{The Belle II Physics Book}},
  \href{https://doi.org/10.1093/ptep/ptz106}{\emph{PTEP} {\bfseries 2019}
  (2019) 123C01} [\href{https://arxiv.org/abs/1808.10567}{{\ttfamily
  1808.10567}}].

\bibitem{Buras:2015qea}
A.~J. Buras, D.~Buttazzo, J.~Girrbach-Noe and R.~Knegjens, \emph{{$ {K}^{+}\to
  {\pi}^{+}\nu \overline{\nu} $ and $ {K}_L\to {\pi}^0\nu \overline{\nu} $ in
  the Standard Model: status and perspectives}},
  \href{https://doi.org/10.1007/JHEP11(2015)033}{\emph{JHEP} {\bfseries 11}
  (2015) 033} [\href{https://arxiv.org/abs/1503.02693}{{\ttfamily
  1503.02693}}].

\bibitem{Martellotti:2015kna}
S.~Martellotti, \emph{{The NA62 Experiment at CERN}},  in \emph{{12th
  Conference on the Intersections of Particle and Nuclear Physics}}, 10, 2015,
  \href{https://arxiv.org/abs/1510.00172}{{\ttfamily 1510.00172}}.

\bibitem{Ertas:2020xcc}
F.~Ertas and F.~Kahlhoefer, \emph{{On the interplay between astrophysical and
  laboratory probes of MeV-scale axion-like particles}},
  \href{https://doi.org/10.1007/JHEP07(2020)050}{\emph{JHEP} {\bfseries 07}
  (2020) 050} [\href{https://arxiv.org/abs/2004.01193}{{\ttfamily
  2004.01193}}].

\bibitem{Wilczek:1977pj}
F.~Wilczek, \emph{{Problem of Strong $P$ and $T$ Invariance in the Presence of
  Instantons}}, \href{https://doi.org/10.1103/PhysRevLett.40.279}{\emph{Phys.
  Rev. Lett.} {\bfseries 40} (1978) 279}.

\bibitem{Haber:1978jt}
H.~Haber, G.~L. Kane and T.~Sterling, \emph{{The Fermion Mass Scale and
  Possible Effects of Higgs Bosons on Experimental Observables}},
  \href{https://doi.org/10.1016/0550-3213(79)90225-6}{\emph{Nucl. Phys. B}
  {\bfseries 161} (1979) 493}.

\bibitem{Haber:1987ua}
H.~E. Haber, A.~S. Schwarz and A.~E. Snyder, \emph{{Hunting the Higgs in $B$
  Decays}}, \href{https://doi.org/10.1016/0550-3213(87)90584-0}{\emph{Nucl.
  Phys. B} {\bfseries 294} (1987) 301}.

\bibitem{Mangano:2007gi}
M.~L. Mangano and P.~Nason, \emph{{Radiative quarkonium decays and the NMSSM
  Higgs interpretation of the hyperCP $\Sigma+ \to p \mu^+ \mu^-$ events}},
  \href{https://doi.org/10.1142/S0217732307023729}{\emph{Mod. Phys. Lett. A}
  {\bfseries 22} (2007) 1373}
  [\href{https://arxiv.org/abs/0704.1719}{{\ttfamily 0704.1719}}].

\bibitem{Domingo:2008rr}
F.~Domingo, U.~Ellwanger, E.~Fullana, C.~Hugonie and M.-A. Sanchis-Lozano,
  \emph{{Radiative Upsilon decays and a light pseudoscalar Higgs in the
  NMSSM}}, \href{https://doi.org/10.1088/1126-6708/2009/01/061}{\emph{JHEP}
  {\bfseries 01} (2009) 061} [\href{https://arxiv.org/abs/0810.4736}{{\ttfamily
  0810.4736}}].

\bibitem{Fayet:2008cn}
P.~Fayet, \emph{{U(1)(A) Symmetry in two-doublet models, U bosons or light
  pseudoscalars, and psi and Upsilon decays}},
  \href{https://doi.org/10.1016/j.physletb.2009.03.078}{\emph{Phys. Lett. B}
  {\bfseries 675} (2009) 267}
  [\href{https://arxiv.org/abs/0812.3980}{{\ttfamily 0812.3980}}].

\bibitem{Merlo:2019anv}
L.~Merlo, F.~Pobbe, S.~Rigolin and O.~Sumensari, \emph{{Revisiting the
  production of ALPs at B-factories}},
  \href{https://doi.org/10.1007/JHEP06(2019)091}{\emph{JHEP} {\bfseries 06}
  (2019) 091} [\href{https://arxiv.org/abs/1905.03259}{{\ttfamily
  1905.03259}}].

\bibitem{Ablikim:2013pqa}
{\scshape BESIII} collaboration, \emph{{Precision measurements of
  B[$\psi$(3686)$\rightarrow$$\pi$+$\pi$-J/$\psi$] and
  B[J/$\psi$$\rightarrow$l+l-]}},
  \href{https://doi.org/10.1103/PhysRevD.88.032007}{\emph{Phys. Rev. D}
  {\bfseries 88} (2013) 032007}
  [\href{https://arxiv.org/abs/1307.1189}{{\ttfamily 1307.1189}}].

\bibitem{Vysotsky:1980cz}
M.~Vysotsky, \emph{{Strong Interaction Corrections to Semiweak Decays:
  Calculation of the V ---\ensuremath{>} H gamma Decay Rate with alpha-S
  Accuracy}}, \href{https://doi.org/10.1016/0370-2693(80)90571-7}{\emph{Phys.
  Lett. B} {\bfseries 97} (1980) 159}.

\bibitem{Nason:1986tr}
P.~Nason, \emph{{QCD Radiative Corrections to $\Upsilon$ Decay Into Scalar Plus
  $\gamma$ and Pseudoscalar Plus $\gamma$}},
  \href{https://doi.org/10.1016/0370-2693(86)90721-5}{\emph{Phys. Lett. B}
  {\bfseries 175} (1986) 223}.

\bibitem{Polchinski:1984ag}
J.~Polchinski, S.~R. Sharpe and T.~Barnes, \emph{{Bound State Effects in
  $\Upsilon \to$ Zeta (8.3) $\gamma$}},
  \href{https://doi.org/10.1016/0370-2693(84)90745-7}{\emph{Phys. Lett. B}
  {\bfseries 148} (1984) 493}.

\bibitem{Pantaleone:1984ug}
J.~T. Pantaleone, M.~E. Peskin and S.~Tye, \emph{{Bound State Effects in
  $\Upsilon \to \gamma$ + Resonance}},
  \href{https://doi.org/10.1016/0370-2693(84)91590-9}{\emph{Phys. Lett. B}
  {\bfseries 149} (1984) 225}.

\bibitem{Aznaurian:1986hi}
I.~Aznaurian, S.~Grigorian and S.~G. Matinyan, \emph{{Relativistic Effects in
  $V \to H^0 \gamma$ Decay}}, {\emph{JETP Lett.} {\bfseries 43} (1986) 646}.

\bibitem{Insler:2010jw}
{\scshape CLEO} collaboration, \emph{{Search for the Decay J/psi -> gamma +
  invisible}}, \href{https://doi.org/10.1103/PhysRevD.81.091101}{\emph{Phys.
  Rev. D} {\bfseries 81} (2010) 091101}
  [\href{https://arxiv.org/abs/1003.0417}{{\ttfamily 1003.0417}}].

\bibitem{Raffelt:1996wa}
G.~G. Raffelt, \emph{{Stars as laboratories for fundamental physics}: {The
  astrophysics of neutrinos, axions, and other weakly interacting particles}}.
  5, 1996.

\bibitem{Bar:2019ifz}
N.~Bar, K.~Blum and G.~D'Amico, \emph{{Is there a supernova bound on axions?}},
  \href{https://doi.org/10.1103/PhysRevD.101.123025}{\emph{Phys. Rev. D}
  {\bfseries 101} (2020) 123025}
  [\href{https://arxiv.org/abs/1907.05020}{{\ttfamily 1907.05020}}].

\bibitem{Chang:2016ntp}
J.~H. Chang, R.~Essig and S.~D. McDermott, \emph{{Revisiting Supernova 1987A
  Constraints on Dark Photons}},
  \href{https://doi.org/10.1007/JHEP01(2017)107}{\emph{JHEP} {\bfseries 01}
  (2017) 107} [\href{https://arxiv.org/abs/1611.03864}{{\ttfamily
  1611.03864}}].

\bibitem{Chang:2018rso}
J.~H. Chang, R.~Essig and S.~D. McDermott, \emph{{Supernova 1987A Constraints
  on Sub-GeV Dark Sectors, Millicharged Particles, the QCD Axion, and an
  Axion-like Particle}},
  \href{https://doi.org/10.1007/JHEP09(2018)051}{\emph{JHEP} {\bfseries 09}
  (2018) 051} [\href{https://arxiv.org/abs/1803.00993}{{\ttfamily
  1803.00993}}].

\bibitem{Raffelt:2006cw}
G.~G. Raffelt, \emph{{Astrophysical axion bounds}},
  \href{https://doi.org/10.1007/978-3-540-73518-2\_3}{\emph{Lect. Notes Phys.}
  {\bfseries 741} (2008) 51}
  [\href{https://arxiv.org/abs/hep-ph/0611350}{{\ttfamily hep-ph/0611350}}].

\bibitem{Keil:1996ju}
W.~Keil, H.-T. Janka, D.~N. Schramm, G.~Sigl, M.~S. Turner and J.~R. Ellis,
  \emph{{A Fresh look at axions and SN-1987A}},
  \href{https://doi.org/10.1103/PhysRevD.56.2419}{\emph{Phys. Rev. D}
  {\bfseries 56} (1997) 2419}
  [\href{https://arxiv.org/abs/astro-ph/9612222}{{\ttfamily
  astro-ph/9612222}}].

\bibitem{Hannestad:1997gc}
S.~Hannestad and G.~Raffelt, \emph{{Supernova neutrino opacity from
  nucleon-nucleon Bremsstrahlung and related processes}},
  \href{https://doi.org/10.1086/306303}{\emph{Astrophys. J.} {\bfseries 507}
  (1998) 339} [\href{https://arxiv.org/abs/astro-ph/9711132}{{\ttfamily
  astro-ph/9711132}}].

\bibitem{Bartl:2016iok}
A.~Bartl, R.~Bollig, H.-T. Janka and A.~Schwenk, \emph{{Impact of
  Nucleon-Nucleon Bremsstrahlung Rates Beyond One-Pion Exchange}},
  \href{https://doi.org/10.1103/PhysRevD.94.083009}{\emph{Phys. Rev. D}
  {\bfseries 94} (2016) 083009}
  [\href{https://arxiv.org/abs/1608.05037}{{\ttfamily 1608.05037}}].

\bibitem{Feng:1997tn}
J.~L. Feng, T.~Moroi, H.~Murayama and E.~Schnapka, \emph{{Third generation
  familons, b factories, and neutrino cosmology}},
  \href{https://doi.org/10.1103/PhysRevD.57.5875}{\emph{Phys. Rev.} {\bfseries
  D57} (1998) 5875} [\href{https://arxiv.org/abs/hep-ph/9709411}{{\ttfamily
  hep-ph/9709411}}].

\bibitem{DEramo:2018vss}
F.~D'Eramo, R.~Z. Ferreira, A.~Notari and J.~L. Bernal, \emph{{Hot Axions and
  the $H_0$ tension}},
  \href{https://doi.org/10.1088/1475-7516/2018/11/014}{\emph{JCAP} {\bfseries
  11} (2018) 014} [\href{https://arxiv.org/abs/1808.07430}{{\ttfamily
  1808.07430}}].

\bibitem{Capozzi:2020cbu}
F.~Capozzi and G.~Raffelt, \emph{{Axion and neutrino red-giant bounds updated
  with geometric distance determinations}},
  \href{https://arxiv.org/abs/2007.03694}{{\ttfamily 2007.03694}}.

\bibitem{Poggio:1975af}
E.~C. Poggio, H.~R. Quinn and S.~Weinberg, \emph{{Smearing the Quark Model}},
  \href{https://doi.org/10.1103/PhysRevD.13.1958}{\emph{Phys. Rev.} {\bfseries
  D13} (1976) 1958}.

\bibitem{Shifman:2000jv}
M.~A. Shifman, \emph{{Quark hadron duality}},  in \emph{{At the frontier of
  particle physics. Handbook of QCD. Vol. 1-3}}, (Singapore), pp.~1447--1494,
  World Scientific, World Scientific, 2001,
  \href{https://arxiv.org/abs/hep-ph/0009131}{{\ttfamily hep-ph/0009131}},
  \href{https://doi.org/10.1142/9789812810458_0032}{DOI}.

\bibitem{Cadamuro:2011fd}
D.~Cadamuro and J.~Redondo, \emph{{Cosmological bounds on pseudo
  Nambu-Goldstone bosons}},
  \href{https://doi.org/10.1088/1475-7516/2012/02/032}{\emph{JCAP} {\bfseries
  02} (2012) 032} [\href{https://arxiv.org/abs/1110.2895}{{\ttfamily
  1110.2895}}].

\bibitem{Millea:2015qra}
M.~Millea, L.~Knox and B.~Fields, \emph{{New Bounds for Axions and Axion-Like
  Particles with keV-GeV Masses}},
  \href{https://doi.org/10.1103/PhysRevD.92.023010}{\emph{Phys. Rev. D}
  {\bfseries 92} (2015) 023010}
  [\href{https://arxiv.org/abs/1501.04097}{{\ttfamily 1501.04097}}].

\bibitem{Depta:2020wmr}
P.~F. Depta, M.~Hufnagel and K.~Schmidt-Hoberg, \emph{{Robust cosmological
  constraints on axion-like particles}},
  \href{https://doi.org/10.1088/1475-7516/2020/05/009}{\emph{JCAP} {\bfseries
  05} (2020) 009} [\href{https://arxiv.org/abs/2002.08370}{{\ttfamily
  2002.08370}}].

\bibitem{NA62:2017rwk}
{\scshape NA62} collaboration, \emph{{The Beam and detector of the NA62
  experiment at CERN}},
  \href{https://doi.org/10.1088/1748-0221/12/05/P05025}{\emph{JINST} {\bfseries
  12} (2017) P05025} [\href{https://arxiv.org/abs/1703.08501}{{\ttfamily
  1703.08501}}].

\bibitem{Alekhin:2015byh}
S.~Alekhin et~al., \emph{{A facility to Search for Hidden Particles at the CERN
  SPS: the SHiP physics case}},
  \href{https://doi.org/10.1088/0034-4885/79/12/124201}{\emph{Rept. Prog.
  Phys.} {\bfseries 79} (2016) 124201}
  [\href{https://arxiv.org/abs/1504.04855}{{\ttfamily 1504.04855}}].

\bibitem{Bergsma:1985qz}
{\scshape CHARM} collaboration, \emph{{Search for Axion Like Particle
  Production in 400-\{GeV\} Proton - Copper Interactions}},
  \href{https://doi.org/10.1016/0370-2693(85)90400-9}{\emph{Phys. Lett. B}
  {\bfseries 157} (1985) 458}.

\bibitem{Dobrich:2017yoq}
{\scshape NA62} collaboration, \emph{{Searches for very weakly-coupled
  particles beyond the Standard Model with NA62}},  in \emph{{13th Patras
  Workshop on Axions, WIMPs and WISPs}}, pp.~145--148, 2018,
  \href{https://arxiv.org/abs/1711.08967}{{\ttfamily 1711.08967}},
  \href{https://doi.org/10.3204/DESY-PROC-2017-02/dobrich\_babette}{DOI}.

\bibitem{Lanfranchi:2017wzl}
{\scshape NA62} collaboration, \emph{{Search for Hidden Sector particles at
  NA62}}, \href{https://doi.org/10.22323/1.314.0301}{\emph{PoS} {\bfseries
  EPS-HEP2017} (2017) 301}.

\bibitem{Ahdida:2654870}
{\scshape SHiP Collaboration} collaboration, \emph{{SHiP Experiment - Progress
  Report}},  Tech. Rep. CERN-SPSC-2019-010. SPSC-SR-248, CERN, Geneva, Jan,
  2019.

\bibitem{CERN-SHiP-NOTE-2015-009}
{\scshape SHiP Collaboration} collaboration, \emph{{Heavy Flavour Cascade
  Production in a Beam Dump}}, .

\bibitem{Clarke:2013aya}
J.~D. Clarke, R.~Foot and R.~R. Volkas, \emph{{Phenomenology of a very light
  scalar (100 MeV \ensuremath{<} $m_h$ \ensuremath{<} 10 GeV) mixing with the
  SM Higgs}}, \href{https://doi.org/10.1007/JHEP02(2014)123}{\emph{JHEP}
  {\bfseries 02} (2014) 123} [\href{https://arxiv.org/abs/1310.8042}{{\ttfamily
  1310.8042}}].

\bibitem{Abt:2007zg}
{\scshape HERA-B} collaboration, \emph{{Measurement of D0, D+, D+(s) and D*+
  Production in Fixed Target 920-GeV Proton-Nucleus Collisions}},
  \href{https://doi.org/10.1140/epjc/s10052-007-0427-z}{\emph{Eur. Phys. J. C}
  {\bfseries 52} (2007) 531} [\href{https://arxiv.org/abs/0708.1443}{{\ttfamily
  0708.1443}}].

\bibitem{Ariga:2018uku}
{\scshape FASER} collaboration, \emph{{FASER's physics reach for long-lived
  particles}}, \href{https://doi.org/10.1103/PhysRevD.99.095011}{\emph{Phys.
  Rev. D} {\bfseries 99} (2019) 095011}
  [\href{https://arxiv.org/abs/1811.12522}{{\ttfamily 1811.12522}}].

\bibitem{Alpigiani:2018fgd}
{\scshape MATHUSLA} collaboration, \emph{{A Letter of Intent for MATHUSLA: A
  Dedicated Displaced Vertex Detector above ATLAS or CMS.}},
  \href{https://arxiv.org/abs/1811.00927}{{\ttfamily 1811.00927}}.

\bibitem{Alpigiani:2020tva}
{\scshape MATHUSLA} collaboration, \emph{{An Update to the Letter of Intent for
  MATHUSLA: Search for Long-Lived Particles at the HL-LHC}},
  \href{https://arxiv.org/abs/2009.01693}{{\ttfamily 2009.01693}}.

\bibitem{Cacciari:1998it}
M.~Cacciari, M.~Greco and P.~Nason, \emph{{The P(T) spectrum in heavy flavor
  hadroproduction}},
  \href{https://doi.org/10.1088/1126-6708/1998/05/007}{\emph{JHEP} {\bfseries
  05} (1998) 007} [\href{https://arxiv.org/abs/hep-ph/9803400}{{\ttfamily
  hep-ph/9803400}}].

\bibitem{Baumholzer:2020hvx}
S.~Baumholzer, V.~Brdar and E.~Morgante, \emph{{Structure Formation Limits on
  Axion-Like Dark Matter}},  \href{https://arxiv.org/abs/2012.09181}{{\ttfamily
  2012.09181}}.

\bibitem{Davidson:1981zd}
A.~Davidson and K.~C. Wali, \emph{{Minimal Flavor Unification via
  Multigenerational Peccei-Quinn Symmetry}},
  \href{https://doi.org/10.1103/PhysRevLett.48.11}{\emph{Phys. Rev. Lett.}
  {\bfseries 48} (1982) 11}.

\bibitem{Wilczek:1982rv}
F.~Wilczek, \emph{{Axions and Family Symmetry Breaking}},
  \href{https://doi.org/10.1103/PhysRevLett.49.1549}{\emph{Phys. Rev. Lett.}
  {\bfseries 49} (1982) 1549}.

\bibitem{Reiss:1982sq}
D.~B. Reiss, \emph{{Can the Family Group Be a Global Symmetry?}},
  \href{https://doi.org/10.1016/0370-2693(82)90647-5}{\emph{Phys. Lett.}
  {\bfseries 115B} (1982) 217}.

\bibitem{Berezhiani:1990wn}
Z.~G. Berezhiani and M.~Y. Khlopov, \emph{{The Theory of broken gauge symmetry
  of families. (In Russian)}}, {\emph{Sov. J. Nucl. Phys.} {\bfseries 51}
  (1990) 739}.

\bibitem{Berezhiani:1990jj}
Z.~G. Berezhiani and M.~Y. Khlopov, \emph{{Physical and astrophysical
  consequences of breaking of the symmetry of families. (In Russian)}},
  {\emph{Sov. J. Nucl. Phys.} {\bfseries 51} (1990) 935}.

\bibitem{Albrecht:2010xh}
M.~E. Albrecht, T.~Feldmann and T.~Mannel, \emph{{Goldstone Bosons in Effective
  Theories with Spontaneously Broken Flavour Symmetry}},
  \href{https://doi.org/10.1007/JHEP10(2010)089}{\emph{JHEP} {\bfseries 10}
  (2010) 089} [\href{https://arxiv.org/abs/1002.4798}{{\ttfamily 1002.4798}}].

\bibitem{Bauer:2016rxs}
M.~Bauer, T.~Schell and T.~Plehn, \emph{{Hunting the Flavon}},
  \href{https://doi.org/10.1103/PhysRevD.94.056003}{\emph{Phys. Rev. D}
  {\bfseries 94} (2016) 056003}
  [\href{https://arxiv.org/abs/1603.06950}{{\ttfamily 1603.06950}}].

\bibitem{Calibbi:2016hwq}
L.~Calibbi, F.~Goertz, D.~Redigolo, R.~Ziegler and J.~Zupan, \emph{{Minimal
  axion model from flavor}},
  \href{https://doi.org/10.1103/PhysRevD.95.095009}{\emph{Phys. Rev.}
  {\bfseries D95} (2017) 095009}
  [\href{https://arxiv.org/abs/1612.08040}{{\ttfamily 1612.08040}}].

\bibitem{Ema:2016ops}
Y.~Ema, K.~Hamaguchi, T.~Moroi and K.~Nakayama, \emph{{Flaxion: a minimal
  extension to solve puzzles in the standard model}},
  \href{https://doi.org/10.1007/JHEP01(2017)096}{\emph{JHEP} {\bfseries 01}
  (2017) 096} [\href{https://arxiv.org/abs/1612.05492}{{\ttfamily
  1612.05492}}].

\bibitem{Ema:2018abj}
Y.~Ema, D.~Hagihara, K.~Hamaguchi, T.~Moroi and K.~Nakayama,
  \emph{{Supersymmetric Flaxion}},
  \href{https://doi.org/10.1007/JHEP04(2018)094}{\emph{JHEP} {\bfseries 04}
  (2018) 094} [\href{https://arxiv.org/abs/1802.07739}{{\ttfamily
  1802.07739}}].

\bibitem{Heikinheimo:2018luc}
M.~Heikinheimo, K.~Huitu, V.~Keus and N.~Koivunen, \emph{{Cosmological
  constraints on light flavons}},
  \href{https://doi.org/10.1007/JHEP06(2019)065}{\emph{JHEP} {\bfseries 06}
  (2019) 065} [\href{https://arxiv.org/abs/1812.10963}{{\ttfamily
  1812.10963}}].

\bibitem{Bonnefoy:2019lsn}
Q.~Bonnefoy, E.~Dudas and S.~Pokorski, \emph{{Chiral Froggatt-Nielsen models,
  gauge anomalies and flavourful axions}},
  \href{https://doi.org/10.1007/JHEP01(2020)191}{\emph{JHEP} {\bfseries 01}
  (2020) 191} [\href{https://arxiv.org/abs/1909.05336}{{\ttfamily
  1909.05336}}].

\bibitem{Egana-Ugrinovic:2019wzj}
D.~Egana-Ugrinovic, S.~Homiller and P.~Meade, \emph{{Light Scalars and the Koto
  Anomaly}}, \href{https://doi.org/10.1103/PhysRevLett.124.191801}{\emph{Phys.
  Rev. Lett.} {\bfseries 124} (2020) 191801}
  [\href{https://arxiv.org/abs/1911.10203}{{\ttfamily 1911.10203}}].

\bibitem{Bonnefoy:2020llz}
Q.~Bonnefoy, P.~Cox, E.~Dudas, T.~Gherghetta and M.~D. Nguyen, \emph{{Flavoured
  Warped Axion}},  \href{https://arxiv.org/abs/2012.09728}{{\ttfamily
  2012.09728}}.

\bibitem{Alonso-Alvarez:2021ett}
G.~Alonso-\'Alvarez, F.~Ertas, J.~Jaeckel, F.~Kahlhoefer and L.~J. Thormaehlen,
  \emph{{Leading Logs in QCD Axion Effective Field Theory}},
  \href{https://arxiv.org/abs/2101.03173}{{\ttfamily 2101.03173}}.

\bibitem{Kaplan:1985dv}
D.~B. Kaplan, \emph{{Opening the Axion Window}},
  \href{https://doi.org/10.1016/0550-3213(85)90319-0}{\emph{Nucl. Phys. B}
  {\bfseries 260} (1985) 215}.

\bibitem{Srednicki:1985xd}
M.~Srednicki, \emph{{Axion Couplings to Matter. 1. CP Conserving Parts}},
  \href{https://doi.org/10.1016/0550-3213(85)90054-9}{\emph{Nucl. Phys. B}
  {\bfseries 260} (1985) 689}.

\bibitem{Chang:1993gm}
S.~Chang and K.~Choi, \emph{{Hadronic axion window and the big bang
  nucleosynthesis}},
  \href{https://doi.org/10.1016/0370-2693(93)90656-3}{\emph{Phys. Lett. B}
  {\bfseries 316} (1993) 51}
  [\href{https://arxiv.org/abs/hep-ph/9306216}{{\ttfamily hep-ph/9306216}}].

\bibitem{Cabibbo:2003cu}
N.~Cabibbo, E.~C. Swallow and R.~Winston, \emph{{Semileptonic hyperon decays}},
  \href{https://doi.org/10.1146/annurev.nucl.53.013103.155258}{\emph{Ann. Rev.
  Nucl. Part. Sci.} {\bfseries 53} (2003) 39}
  [\href{https://arxiv.org/abs/hep-ph/0307298}{{\ttfamily hep-ph/0307298}}].

\bibitem{Jaffe:1989jz}
R.~L. Jaffe and A.~Manohar, \emph{{The G(1) Problem: Fact and Fantasy on the
  Spin of the Proton}},
  \href{https://doi.org/10.1016/0550-3213(90)90506-9}{\emph{Nucl. Phys. B}
  {\bfseries 337} (1990) 509}.

\end{thebibliography}\endgroup
\bibliographystyle{JHEP}

\end{document}